\documentstyle[emulateapj,psfig,apjfonts]{article}
\submitted{Received 2002 May 7; Accepted: 2002 August 27}

\def\gs{\mathrel{\raise0.35ex\hbox{$\scriptstyle >$}\kern-0.6em
\lower0.40ex\hbox{{$\scriptstyle \sim$}}}}
\def\ls{\mathrel{\raise0.35ex\hbox{$\scriptstyle <$}\kern-0.6em
\lower0.40ex\hbox{{$\scriptstyle \sim$}}}}

\lefthead{Smail et al.}
\righthead{The Properties of $\mu$Jy EROs}

\makeatletter

\newenvironment{inlinefigure}{%
\def\@captype{figure}%
\noindent\begin{minipage}{0.999\linewidth}\begin{center}\small}
{\end{center}\end{minipage}\smallskip}
\makeatother

\begin{document}

\title{The Diversity of Extremely Red Objects}

\author{
Ian Smail,\altaffilmark{1,2}
F.\,N.\ Owen,\altaffilmark{2,3} G.\,E.\ Morrison,\altaffilmark{2,4}
W.\,C.\ Keel,\altaffilmark{2,5} R.\,J.\ Ivison\altaffilmark{6} \&
M.\,J.\ Ledlow\altaffilmark{2,7}
}
\altaffiltext{1}{Insitute for Computational Cosmology, 
University of Durham, South Road, Durham DH1 3LE}
\altaffiltext{2}{Visiting astronomer, Kitt Peak National Observatory,
National Optical Astronomy Observatories, operated by AURA, Inc.,
under cooperative agreement with the National Science Foundation.}
\altaffiltext{3}{National Radio Astronomy Observatory, P.\ O.\ Box O,
Socorro, NM 87801 USA}
\altaffiltext{4}{California Institute of Technology, IPAC, MS\,100-22,
Pasadena, CA 91125 USA}
\altaffiltext{5}{Dept. of Physics \& Astronomy, University of Alabama,
Tuscaloosa, AL 35487 USA}
\altaffiltext{6}{Astronomy Technology Centre, Royal Observatory, 
Blackford Hill, Edinburgh EH9 3HJ}
\altaffiltext{7}{Gemini Observatory, Southern Operations Center, AURA,
Casilla 603, La Serena, Chile}

\setcounter{footnote}{8}

\begin{abstract}
We present the results from a sensitive multi-wavelength analysis of
the properties of Extremely Red Objects (EROs).  Our analysis employs
deep $RIzJHK$ photometry of a $8.5'\times 8.5'$ region to select a
sample of 68 EROs with $(R-K)\geq 5.3$ and brighter than $K=20.5$
($5\sigma$).  We combine this photometric dataset with an extremely
deep 1.4-GHz radio map of the field obtained from the VLA.  This map
reaches a 1-$\sigma$ limiting flux density of 3.5$\mu$Jy making it the
deepest 1.4-GHz map taken and is sensitive enough to detect an active
galaxy with $L_{1.4}\gs 10^{23}$\,W\,Hz$^{-1}$ at $z>1$. If powered by
a starburst, this radio luminosity is equivalent to a star-formation
rate of $\gs 25$\,M$_\odot$\,yr$^{-1}$ for stars more massive than
5\,M$_\odot$.  We identify radio counterparts to 21 of the EROs in this
field with radio fluxes above 12.6$\mu$Jy and resolve a third of these
with our 1.6$''$ FWHM beam. The spectral energy distributions of the
majority of these galaxies are consistent with those expected for
dust-reddened starbursts at $z\sim 1$.  At these redshifts the radio
luminosities of these galaxies indicate a median far-infrared
luminosity of this population of $L_{FIR}\gs 10^{12}L_\odot$, meaning
half of the radio-detected sample are ultraluminous infrared galaxies
(ULIRGs).  We conclude that $\gs 16\pm 5$\% of the ERO population
brighter than $K=20.5$ are luminous infrared galaxies (LIRGs) at $z\sim
1$. We also use photometric classification of the colors of the EROs to
investigate the mix of dusty active and evolved passive systems in the
remaining ERO population which is undetected in our radio map. Based on
this we suggest that at least 30\% and possibly up to $\sim 60$\% of
{\it all} EROs with $(R-K)\geq 5.3$ and $K\leq 20.5$ are dusty,
star-forming systems at $z\gs 1$.  Our best estimate of the star
formation density in this highly-obscured and optically faint ($R\gs
26$) population is $\stackrel{.}{\rho_*}(0.1$--100\,M$_\odot)=0.11\pm
0.03$\,M$_\odot$\,yr$^{-1}$\,Mpc$^{-3}$, comparable to estimates of
that in H$\alpha$ emitting galaxies at $z\sim1$, and greater than the
estimates from UV-selected samples at these epochs.  This lends support
to the claims of a strong increase in the contribution from obscured
systems to the star formation density at high redshifts.  Using the
observed counts of the radio-detected ERO population we model the
apparent break in the $K$-band number counts of the whole ERO
population at $K\sim 19$--20 and propose that the passive ERO class
dominates the total population in a relatively narrow magnitude range
around $K\ls 20$, with dusty, active EROs making up the bulk of the
population at fainter limits.
\end{abstract}


\section{Introduction}

The last five years have seen a growing appreciation of the diversity
of galaxy properties at $z\gs 1$--3.  In part this has arisen from an
impressive improvement in the quality of the observations of galaxies
at these redshifts -- driven by the availability of powerful new
instruments on 8-m class telescopes (e.g.\ NIRSPEC on Keck and ISAAC on
the VLT).  However, an equal role has been played by the realization of
the necessity of a multi-wavelength approach to studies of galaxy
evolution at $z\gs 1$.  Thus a range of new surveys spanning wavebands
from the X-ray, near- and mid-infrared, submillimeter and out to the
radio, have complimented the traditional view of distant galaxies based
on UV/optical observations.  These new studies have all tended to
stress the role of dust obscuration in censoring our view of the galaxy
population at high redshifts, and especially in disguising the extent
of activity in the most luminous systems, both AGN and star-forming.
Indeed, our current, incomplete, knowledge of the evolution of dusty
galaxies suggests that some of the most active galaxies in the distant
Universe may be so obscured as to be undetectable in even the most
sensitive UV/optical observations (e.g.\ Haarsma et al.\ 2000; Smail et
al.\ 2002).

One central theme to appear from this new multi-wavelength view of
galaxy formation is the ubiquity of optically-faint, but bright
near-infrared, counterparts to sources identified in many wavebands: in
the X-ray (Alexander et al.\ 2001, 2002; Cowie et al.\ 2001; Page et
al.\ 2002; Stevens et al.\ 2002; Mainieri et al.\ 2002), the
mid-infrared (Pierre et al.\ 2001; Smith et al.\ 2001; Franceschini et
al.\ 2002), submillimeter (Smail et al.\ 1999a; Lutz et al.\ 2001;
Ivison et al.\ 2002) and radio bands (Richards et al.\ 1999; Chapman et
al.\ 2001).  This has resulted in a renaissance in interest in such
Extremely Red Objects (EROs) -- a class of galaxies which had
previously been viewed as a curiosity with little relevance to our
understanding of galaxy formation and evolution, comprising as they do
a mere $\sim 5$--10\% of the population at $K\leq 20$.

The ERO class is photometrically-defined -- the most frequently used
definition is now $(R-K)\geq 5.3$.  This very red
optical-near-infrared color is intended to isolate two broad
populations of galaxies at $z\gs 1$: those which are red by virtue of
the presence of large amounts of dust (resulting from active star
formation), as well as passive systems whose red colors arise from the
dominance of old, evolved stars in their stellar populations.  The
very different natures of these two sub-classes has prompted efforts
to disentangle their relative contributions to the ERO population
(Pozzetti \& Mannucci 2000; Mannucci et al.\ 2002) so that
well-defined samples can be used to test galaxy formation models
(Daddi et al.\ 2000; Smith et al.\ 2002a; Firth et al.\ 2002).

Studies of the ERO population have turned up hints of the diverse
nature of this population, including an apparent break in the
number counts of EROs at $K\sim 19$--20 (McCarthy et al.\ 2001; Smith
et al.\ 2002a).  This feature is roughly 2\,mags fainter than the well
known break in the total $K$-band counts of the faint field population
(Gardner et al.\ 1993). The cumulative count slope of EROs
($\log_{10}N(<K) \propto \alpha K$) drops from $\alpha= 0.81\pm 0.05$
at bright magnitudes to $\alpha= 0.32\pm 0.05$ for the fainter EROs
(e.g.\ Smith et al.\ 2002a; Chen et al.\ 2002; Fig.~2).  This break in
the counts could either result from a relatively narrow redshift
range probed by the selection criteria of EROs -- with the change in
the integrated counts then being simply related to the form of the
luminosity function of the ERO population, or it may reflect differing
evolutionary behavior in sub-populations -- where one class of EROs
declines rapidly beyond $K\sim 19$.  To test this suggestion we need
to track the relative mix of galaxy types within the ERO population to
$K>19$--20.

Techniques to separate the passive, evolved and obscured, active
populations include near-infrared photometric classifications (Pozzetti
\& Mannucci 2000; Smith et al.\ 2002a; Mannucci et al.\ 2002),
submillimeter emission (Dey et al.\ 1999; Mohan et al.\ 2002), X-ray
emission (Alexander et al.\ 2001, 2002; Brusa et al.\ 2002) and
optical/near-infrared spectroscopy (Cimatti et al.\ 2002; Smith et al.\
2001, 2002b).  Of these approaches, by far the largest and most
influential project is the ``K20'' survey undertaken by Cimatti et al.\
(2002) who obtained spectroscopic classifications for 29 EROs with
$K<19.2$ and $(R-K)\geq 5$ from a full sample of 45. The median
redshift of their $K<19.2$ sample is $z=1.1\pm 0.2$, with galaxies
spread across $z=0.7$--1.4 (see Daddi et al.\ 2001) and they find that
their sample is split equally into passive, evolved and active, dusty
EROs: 31--64\% versus 67--33\%, where the ranges reflect the
uncertainty due to the incompleteness in their survey.  The K20 survey
has significantly increased our knowledge of the ERO population,
however, it is reliant on identifying emission lines in the restframe
UV spectra of possibly dusty galaxies, which results in an incomplete
and perhaps biased view of the mix of galaxies within the population.
Moreover, the magnitude limit achieved for the high completeness
sample, $K<19.2$, means that it can't be used to track the variation in
the properties of the population across the break in the counts of EROs.

Here we apply a new technique to investigate the mix of the ERO
population at faint magnitudes: using very deep radio observations at
1.4\,GHz which are both sensitive enough to identify strongly star
forming galaxies out to high redshifts, $z\gs 1$--2, and yet
insensitive to the effects of dust obscuration (Mohan et al.\ 2002;
Chapman et al.\ 2002a).  Radio surveys to sub-mJy flux limits have
shown the usefulness of this technique for identifying star-forming
galaxies in the distant Universe -- morphological classifications from
{\it Hubble Space Telescope} ({\it HST}) imaging show many are blue
disk galaxies (Windhorst et al.\ 1994; Richards et al.\ 1998; Richards
2000), with optical spectroscopy confirming that these are apparently
normal star forming galaxies at intermediate redshifts (Benn et al.\
1993; Mobasher et al.\ 1999; Roche, Lowenthal \& Koo 2002), with a small
proportion of AGN (20\% based on their radio morphologies in very
high-resolution maps from combined VLA and Merlin observations, Muxlow
et al.\ 1999, with most of these having blue colors).  However, the
spectral properties of the galaxies in these surveys also show
signatures of the presence of dust, which may be cloaking the strength
of the activity in these systems and making them appear more mundane
than they should (Hammer et al.\ 1995; Poggianti \& Wu 2000; Smail et
al.\ 1999b).

More recently, samples of faint radio sources have begun to be used to
attempt to identify obscured star-forming galaxies at even higher
redshifts. By focusing on those radio sources with faint or
undetectable counterparts in the optical these surveys attempt to
isolate the most distant star forming galaxies in the radio samples
(Richards et al.\ 1999; Barger et al.\ 2000; Chapman et al.\ 2001,
2002b).  Sensitive submillimeter observations of these optically-faint
radio sources confirm that a large fraction of them are luminous, dusty
systems at redshifts of $z\sim 1$--3 -- although, as with the whole
submillimeter galaxy population, there is still considerable
uncertainty over their exact redshifts (Chapman et al.\ 2002b).  The
relationship between these highly obscured, but very luminous galaxies
and the perhaps less-obscured systems selected through Lyman-$\alpha$
emission and the Lyman-break technique is a crucial issue for a
complete understanding of the formation and evolution of galaxies
(Smail et al.\ 2002, 2003).  The project described here seeks to trace
the population of the luminous, dusty galaxies detected in the
submillimeter at $z\gs 2$ to lower redshifts and lower luminosities.
These samples should be more amenable to detailed study and so provide
a clearer connection between the evolution of the obscured and
unobscured galaxies over the lifetime of the Universe (Chapman et al.\
2002b).

In this paper we use an extremely deep VLA radio map to investigate
the radio emission from a sample of ERO galaxies at a sensitivity
limit sufficient to detect strongly star forming galaxies at $z\gs
1$--2.  We also present extensive optical/near-infrared imaging
of this ERO sample.  We exploit this deep $RIzJHK$ photometry to
analyse the spectral energy distributions (SEDs) of the EROs and
investigate their properties in detail.  We present our observations,
their reduction and cataloging in the next section, describe our
analysis and results in \S3, and discuss these in \S4 before giving
our conclusions in \S5.  Throughout we assume a cosmological model
with $q_o=0.5$ and $H_o=50$\,km\,s$^{-1}$\,Mpc$^{-1}$.  If we instead
adopted the currently fashionable $H_o=70$\,km\,s$^{-1}$\,Mpc$^{-1}$,
$\Omega_\Lambda=0.7$, $\Omega_m=0.3$ cosmology, then for a source at
$z=1$ all the linear sizes we derive would be 6\% smaller, the
luminosities would be 12\% fainter and the volume densities
would be 12\% lower.

\section{Observations, Reduction and Cataloging}

The field used for our analysis contains the $z=0.41$ cluster A\,851
(Cl\,0939+4713). This well-studied cluster is relatively rich, but
very irregular in appearance (Dressler et al.\ 1994, 1997, 2002; Iye
et al.\ 2001; Schindler et al.\ 1998).  As we discuss below, although
cluster members may contribute significantly to the optically-bright
radio population in this field, the extremely red radio sources
discussed in this work all probably lie at $z\gs 1$ and hence are
unassociated with the foreground cluster.

The other influence which the cluster may have is as a gravitational
lens. However, A\,851 is not a particularly concentrated cluster
(Seitz et al.\ 1996; Trager et al.\ 1997; Iye et al.\ 2000) and the
wide field coverage of our sample, compared with the cluster's
critical radius (only two of the EROs in our sample lie within 1$'$ of
the cluster center, roughly corresponding to 2--$3\times$ the critical
radius), suggests that lensing by the cluster will have a minimal
effect on the surface density in our sample ($\ls 5$\% integrated over
the whole area of our survey) and is unlikely to grossly effect the
morphologies of any of the EROs in our study.

\subsection{Radio Observations}

The radio image of A\,851 comprises a combination of A, B, C, and D
configuration observations from the National Radio Astronomy
Observatory's VLA\footnote{NRAO is operated by Associated Universities
Inc., under a cooperative agreement with the National Science
Foundation.} taken at 1.4\,GHz between 1996 and 2000. For each
observation, ``4'' mode was used for the correlator with $7\times
3.125$ MHz spectral channels for each of two IFs and two
polarizations. This mode was employed to allow a map to be made of the
entire primary beam at the highest spatial resolution possible with
the VLA ($\sim 1.5''$ FWHM) and with the best compromise between
radial smearing of the image due to the finite bandwidth and
sensitivity.  However, as is noted below, some penalty does result
from this choice. The total integration time of the combined dataset
is $\sim 100$ hours on source.

The dataset was imaged using the {\sc aips} program {\sc imagr}. The
74 facets were used including 37 within the primary beam and 37
outside the primary beam, each of the latter centered on a bright
confusing source.  Each of the datasets from the four configurations
were self-calibrated using intermediate images as input to the {\sc
aips} program {\sc calib}. A more complete description of the
reduction process is included in Owen et al.\ (2002).

The analysis presented here focuses on the central facet from the full
map, a $8.53'\times 8.53'$ field in the center of the primary beam: 09
42 48.614 +46 59 59.72 (J2000). The final map reaches a 1-$\sigma$
noise level of 3.5\,$\mu$Jy\,beam$^{-1}$ in the central part of this
field -- making this the deepest 1.4-GHz map ever made.  However, in
the extreme corners of the area being analyzed the actual peak flux
density is reduced by about $20\%$ due to the primary beam attenuation
and radial bandwidth smearing caused by the finite bandwidth of each
spectral channel.  We therefore adopt a conservative 3-$\sigma$ limit
of 12.6\,$\mu$Jy in the corrected peak flux density over the entire
area being studied.  We can probe down to 3-$\sigma$ significance limit
and still retain a very low level of false detections as we are
searching for radio emission at the known positions of a relatively
small number of EROs, compared to the number of independent beams in
our map ($10^5$).  The synthesized beam of our final map is circular
and has a FWHM of $1.6''$, this allows us to resolve radio sources with
sizes $\gs 1''$ and we discuss the sizes of radio emission from the
EROs in our survey in the following sections.

1%
%
\begin{figure*}[tbh]
\centerline{\psfig{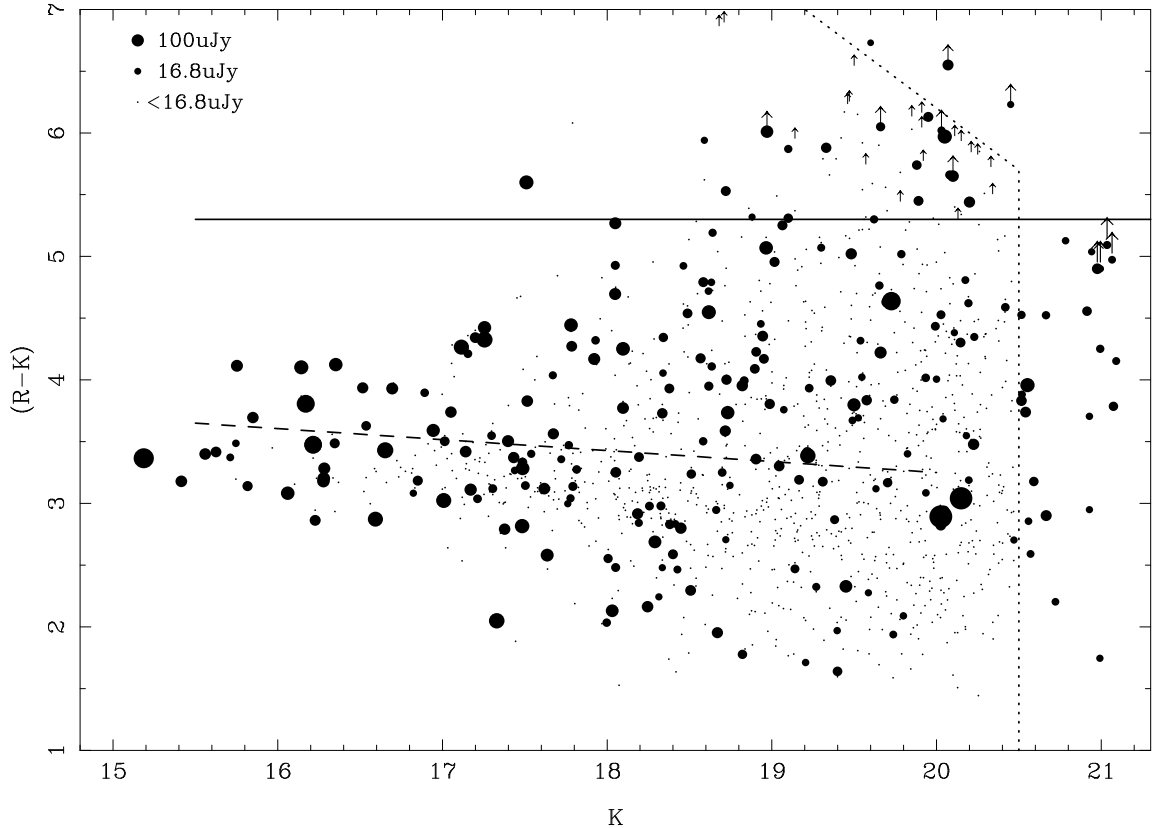}}
\caption{\small The $(R-K)$--$K$ color-magnitude diagram for 1186
sources brighter than $K=20.5$ within a $8.53'\times 8.53'$ field
centered on A\,851 -- these are shown as small points and lower
limits.  Overlayed on this are the radio-detected sources brighter
than the 4-$\sigma$ limit of our radio map, 16.8\,$\mu$Jy.  These
radio sources go fainter than the nominal $K=20.5$ catalog limit
(shown by the dotted line). The solid line shows the $(R-K)=5.3$
boundary used to define our ERO sample. For the ERO sample we identify
all sources detected above a slightly deeper radio limit,
12.7\,$\mu$Jy or $3\sigma$, again indicating their radio fluxes by the
size of the symbols used.  The dashed line denotes the reddest color
expected for unobscured galaxies at the cluster redshift; galaxies
redder than this limit are likely to be beyond the cluster.  The
population of radio-emitting cluster members is clearly visible below
this boundary (Morrison 1999; Morrison et al.\ 2002). }
\end{figure*}

\subsection{Near-infrared Observations}

Our primary near-infrared imaging dataset comes from wide-field images
in $JHK$ taken with the KPNO 2.1-m telescope, these cover a
$15.4'\times 15.7'$ field of view to $K\sim 21$. These images have
relatively coarse sampling and so we have also obtained higher
resolution $J$ and $K$ images covering a smaller, $7.5'\times 7.5'$,
field of view from the 4.2-m William Herschel Telescope
(WHT)\footnote{The WHT is operated by the Isaac Newton Group on behalf
of Particle Physics \& Astronomy Research Council (PPARC).} and the
3.8-m UK Infrared Telescope (UKIRT)\footnote{UKIRT is operated by the
Joint Astronomy Centre on behalf of PPARC.} respectively. The WHT and
UKIRT observations cover the bulk of the region included in the
analysis here and they thus provide a useful confirmation of the
identifications and photometry of EROs in this field.  Standard
reduction procedures were applied to all datasets and they were
calibrated using UKIRT Faint Standards (Hawarden et al.\ 2001).

\subsubsection{KPNO 2.1\,m $JHK$-band mosaics}

The SQIID camera was used at the KPNO 2.1-m telescope on the nights of
2001 January 4--11 to image the field simultaneously in $JHK$.  The
pixel scale of the camera is $0.68$ arcsec and the useful field of view of
is $\sim 5.2'\times5.2'$, with nine pointings mosaiced together to cover a
region of $15.4'\times 15.7'$.  The total exposure time was 75\,ks,
spread equally over the nine pointings to give a per-pixel exposure
time of 8.3\,ks.

The images were reduced in a standard manner using {\sc iraf} scripts.
Flat fields were constructed from the dithered exposures at each
pointing to produce a flat field centered in time on each exposure.
Distortion corrections were measured from a comparison of bright
objects in the near-infrared images with existing, deep $R$-band
images and these were applied to align the images in the three
passbands to a common coordinate grid, at a level of 0.1 pixel rms or
better.

These observations were obtained in a mix of photometric and
non-photometric conditions, with seeing of 1.5--1.8$''$ FWHM.  To
calibrate the photometry in the three bands we exploit the precise
$JHK$ photometry of this field published by Stanford, Eisenhardt \&
Dickinson (1995). By comparing the photometry for a large number of
bright galaxies in the field we estimate a typical zero-point error of
$\ls 0.03$\,mags on our photometric scale.  The total magnitude
corresponding to the 5-$\sigma$ detection limit for a point source is
$K=20.5$. While the 3-$\sigma$ limits for 3$''$ aperture photometry in
the $J$- and $H$-bands are $J=22.7$ and $H=21.9$.

\subsubsection{UKIRT $K$-band mosaic}

On the nights of 1999 February 28, March 01--02 and March 15--16 we
completed a $K$-band mosaic of $5\times 5$ pointings using the UFTI
imager on UKIRT (pixel scale 0.0908$''$ pixel$^{-1}$).  This mosaic
covers the central 7.5$'\times 7.5'$ region of the cluster defined by
the {\it WFPC2} field (see \S2.3) and hence covers around 55\% of the
radio sample analysed here. Each pointing has a total integration time
of 2.4\,ks and reaches a limit of $K\sim 20.0$ for point sources. The
median seeing for these observations was 0.5$''$ and several of the
mosaic tiles have 0.35$''$ seeing, while periods of sub-0.25$''$
seeing were experienced.  These frames allow us to confirm the reality
of EROs in the field and investigate the morphologies of the brighter
examples.

\subsubsection{WHT $J$-band mosaic}

On 2000 December 11--12 we used the INGRID near-infrared imager
(Packham et al.\ 2002) on the WHT to obtain a $J$-band image of the
cluster core. This comprised a $2\times 2$ mosaic of 2.4-ks exposures
and covers the same region as the UKIRT mosaic, reaching $J=22.7$ for
a point-source in 0.70$''$ seeing with 0.242$''$ sampling.

\subsection{Optical Observations}

The optical imaging to complement our deep near-infrared datasets was
obtained from three facilities.  We have two wide-field imaging
datasets of this region: the first comes from Suprime-Cam on Subaru
(courtesy of Dr T.\ Kodama), and consists of very deep $RI$ imaging
of a $27'\times 27'$ field centered on the cluster core. We also have
obtained similarly wide-field KPNO 4-m image in the $z$-band 
to bridge the spectral coverage of our optical and near-infrared
datasets.  Finally, an {\it HST/WFPC2} mosaic of A\,851 has been
observed in the F702W filter, this covers a 7.5$'\times 7.5'$ region
around the cluster core, within the area studied here, and provides
very detailed morphological information on the optically-bright
galaxies in this area (Dressler et al.\ 2002; Morrison et al.\
2002). However, the relatively shallow surface brightness limit of the
{\it HST} imaging, $\mu_R\sim 25.5$\,mag\,arcsec$^{-2}$, means that
these data are not useful for gauging the morphologies of all but the
most compact of the 15 radio-detected and 27 radio-undetected EROs in
this region (see also Dey et al.\ 1999; Smith et al.\ 2002a). For this
reason we will not use the {\it HST} imaging in our analysis.

\subsubsection{Subaru $RI$ imaging}

Kodama et al.\ (2001) obtained deep, $RI$ imaging of A\,851 using
Suprime-Cam on the 8.3-m Subaru Telescope on the nights of 2001
January 21--22. Suprime-Cam has a $27'\times 27'$ field of view,
imaged with eight 4k$\times$2k CCDs at a scale of 0.20$''$ pixel$^{-1}$. The
conditions were photometric during the observations, with seeing of
1.0$''$ and 0.7$''$ in $R$ and $I$ respectively.  Total exposure times
of 4.0\,ks and 1.3\,ks were obtained resulting in 3-$\sigma$ limiting
magnitudes for point sources in 3$''$ diameter apertures of $R=26.7$
and $I=25.6$. The reduction and calibration of these data is described
by Kodama et al.\ (2001).

\subsubsection{KPNO 4\,m $z$ imaging}

The $z$-band imaging of A\,851 was obtained on the night of 2001
January 25--26 with the MOSAIC camera on the KPNO 4-m. These data
cover a 36$'\times 36'$ field with 0.258$''$ pixel$^{-1}$ sampling.
The processed image combined five individual exposures, shifted so the
combination eliminated areas lost to gaps between individual CCD
chips. After applying chip-by-chip sky flats, each exposure was
projected onto an astrometric pixel grid and coadded with appropriate
masking for gaps and bad pixels.  The data were taken in photometric
conditions and the final 7.2\,ks exposure has seeing of 1.0$''$ FWHM
and a 3-$\sigma$ limiting magnitude of $z=24.1$ in our standard 3$''$
diameter aperture.

\subsection{Photometric Catalog}

To catalog near-infrared sources in the field of A\,851 we ran
SExtractor (Bertin \& Arnouts 1996) on the $K$-band image of the field
-- first convolving the image with a $2''$-diameter top-hat kernel and
then searching for sources with areas greater than the seeing-disk
above an isophote defined as $\mu_K=22.8$ mag.\ arcsec$^{-2}$.

As our measure of the total magnitudes of the sources in the image we
adopt the scaled aperture magnitude, {\sc best\_mag}, calculated by
SExtractor.  This magnitude is measured within an aperture which is
$2.5\times$ the first-moment radius of the source, except in crowded
environments when it is replaced with a corrected, isophotal magnitude,
see Bertin \& Arnouts (1996) for details.  The 5-$\sigma$ detection
limit for point-sources in our catalog corresponds to a total magnitude
of $K=20.5$ and we detect 1186 sources in the area spanned by our
current VLA catalog (Fig.~1).  This point-source detection limit should
be representative of the completeness of our catalog for those compact
galaxies with half-light radii less than our $2''$-diameter detection
filter (this includes the majority of the faint, near-infrared field
population, Chen et al.\ 2002).  However, our catalog selection will
tend to miss galaxies with very extended disks even though their
integrated magnitudes are above this point-source limit and this may
introduce a morphological bias into our sample selection (Martini 2001;
Chen et al.\ 2002; Firth et al.\ 2002; Snigula et al.\ 2002).  However,
this is not a significant problem for our analysis as the primary goal
is to determine the proportion of dusty and evolved EROs in published
surveys and the relatively good agreement between our measured ERO
surface density and values published by other workers (see Fig.~2)
suggests that our catalog suffers no stronger bias than is typical for
such surveys.

To measure colors for galaxies in our $RIzJHK$ imaging we use fixed,
$3''$-diameter aperture photometry from seeing-matched frames. The
optical and near-infrared images were aligned to the coordinate system
defined by the KPNO $K$-band image with a typical tolerance of $\ls
0.1''$.  The seeing on these frames was then matched to that of the
poorest-resolution images (the $J/H/K$-band) by Gaussian convolution.

\section{Analysis and Results}

Our analysis begins by constructing a SExtractor-based $K$-selected
catalog of EROs in the field.  We then align the radio and
optical/near-infrared images to an rms accuracy of $0.32''$ using the
90 brightest radio and $K$-band sources in the joint field. This
allows us to search for matches for the EROs within the radio map.  We
note that our radio observations identify all three cataloged submm
galaxies in this field (Smail et al.\ 1999a, 2002; Cowie et al.\ 2002)
and we discuss these further in Ledlow et al.\ (2002).

\subsection{The ERO samples}

For this analysis we start from our $K$-selected galaxy catalog and
identify galaxies brighter than $K=20.5$ ($5\sigma$) which have 3-$''$
diameter aperture colors redder than the limit of $(R-K)=5.3$ which we
adopt as the definition of an ERO (see Pozzetti \& Mannucci 2000).
Based on this we identify 68 EROs in our field -- equivalent to a
surface density of $(1.04\pm0.12)$ arcmin$^{-2}$ at $K=20.5$ (the
error-bar simply reflects Poisson statistics and we have applied an
incompleteness correction of 20\% to the number density of EROs in the
faintest bin, as well as a 6\% correction to the area of the survey to
account for the sky area covered by bright stars and galaxies).

The catalog of EROs in our field is given in Table~1.  This lists the
ERO ID (a 5-digit ID denotes a radio source), the J2000 position of
the ERO (tied to the VLA reference frame with an rms precision of
0.32$''$), the total $K$-band magnitude (measured using the SExtractor
{\sc best\_mag}), aperture $(R-K)$ color (or 3-$\sigma$ limit),
1.4-GHz radio flux (or 3-$\sigma$ limit), 3-$''$ diameter aperture
magnitudes from our seeing-matched $RIzJHK$ imaging, best-fitting SED
(either dusty, ``D''; evolved, ``E''; or badly-fit, ``B'', see \S3.5)
and a visual morphology (from our high resolution $K$- and $R$-band
imaging if available, see \S3.4) on a simple system: ``C'',
compact/regular; ``I'', disky, disturbed or amorphous; ``F'', too
faint to classify.

%
%
\smallskip
\begin{inlinefigure}
\centerline{\psfig{file=f2.ps,width=3.3in} }
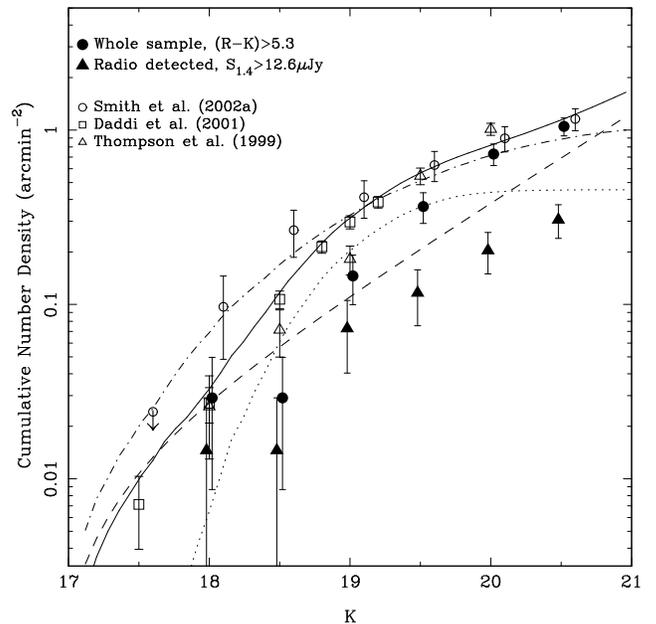
\caption{\small The cumulative number density of EROs in our sample
compared to published counts from Daddi et al.\ (2000), Thompson et
al.\ (1999) and Smith et al.\ (2002a). We show both the full ERO
catalog, selected with $(R-K)>5.3$ and also the radio-detected
sample. The error-bars represent simple Poisson uncertainties and are
likely therefore to be underestimates of the true field-to-field
variations (Daddi et al.\ 2001). We have included a 6\% downward
correction to the estimated area of our survey due to coverage by
bright stars and galaxies in our field.  The dashed line shows the fit
to the radio-detected ERO population with a normalization increased to
reproduce the 60\% dusty ERO fraction estimated for our whole sample to
$K=20.5$.  The dotted line is a Gaussian component with a mean
magnitude of $K=19.1$ and a FWHM of 1.2\,mags which is normalized so
as to reproduce the cumulative counts of EROs in our field.  The solid
line shows the behavior of a toy model constructed by combining the
passive and dusty populations indicated by the dashed and dotted
lines.  The dot-dashed line shows a second toy model which combines a
series of Gaussian luminosity functions for the ERO population with a
mean of $L_K^\ast$ and a dispersion of $\sigma_K=1$\,mag, which are
coadded so as to produce a Gaussian redshift distribution with $<\!
z\! >\sim 1.1$ and $\sigma_z=0.2$ (see Firth et al.\ 2002).  }
\end{inlinefigure}

As is shown in Fig.~2, the surface density of EROs we detect is in
reasonable agreement with that observed in other ERO surveys
(e.g.\ Thompson et al.\ 1999; Daddi et al.\ 2000; Smith et al.\ 2002a)
given the strong field-to-field variation seen in the surface density
of EROs (Daddi et al.\ 2001; McCarthy et al.\ 2001; Roche et al.\
2002; Chapman, McCarthy \& Persson 2000). This supports our claim that
the foreground cluster makes no contribution to this population, either
from members or by gravitational lensing.  For
ease of comparison with ERO samples selected in other filter
combinations, in particular $(I-K)$, we note that while our field
contains 68 EROs with $(R-K)\geq 5.3$, we have 99 objects with
$(I-K)\geq 4$ suggesting that this is a less stringent definition of
an unusually red source. We also note that 63 (93\%) of our $(R-K)\geq
5.3$ EROs have $(I-K)\geq 4$ (including all of the radio-detected EROs
from \S3.2).

We re-identify three previously cataloged EROs brighter than $K=20.5$
in our field, including the submm ERO SMM\,J09429+4658 from Smail et
al.\ (1999a), which is also a faint radio source, and two other
radio-undetected EROs -- the edge-on disk galaxy \#333 from the {\it
NICMOS} study of Smail et al.\ (1999b) and Cl\,0939+4713 A from
Persson et al.\ (1993). A fourth cataloged ERO in this field -- the
passive $z=1.58$ ERO Cl\,0939+4713 B from Soifer et al.\ (1999) -- has
$(R-K)=5.1\pm 0.1$ based on our photometry and hence does not make it
into our ERO sample.

We can also use our UKIRT and WHT near-infrared imaging to test the
reliability of our ERO sample.  There are 30 EROs from the KPNO sample
which fall within the UKIRT $K$-band and WHT $J$-band mosaics (11 of
these are radio-detected).  We identify 27 (90\%) of these on the
UKIRT mosaic. Two of the missing EROs are low-surface brightness,
extended systems which are below the detection limit of the UKIRT
mosaic when using a large diameter photometric aperture.  This
highlights the surface brightness, and hence morphological, biases
which could affect the mix of morphological classes identified in ERO
surveys (Chen et al.\ 2002).  In this regard the good sensitivity, but
relatively coarse, resolution of our KPNO $K$-band imaging acts in our
favor by concentrating a large fraction of the light from the galaxy
in a relatively small number of pixels ($50\times$ fewer than in the
UKIRT image).

The third ``ERO'' missing from the UKIRT mosaic is the source \#2549
cataloged in the KPNO 2.1-m imaging (Table~1).  This source is clearly
detected in the simultaneously obtained $JHK$ imaging from 2001 January
4--11 as a bright $K=19.0$ unresolved source ($<1.5''$ FWHM) with
unusual colors $(J-K)=0.22\pm 0.07$, $(H-K)=-1.09\pm 0.07$: indicating
that the SED peaks in the $H$-band (Fig.~7). Yet it was apparently
fainter than $K>20.0$ during February 1999 (UKIRT) and $J>22.7$ in
December 2000 (WHT).  In addition, we can state the counterpart to this
source was also fainter than our 3-$\sigma$ limits of $R>26.7$ and
$I>25.6$ on 2001 January 21 and $z>24.1$ on 2001 January 25.  The
strongest constraint on the variability of the source comes from our
$J$-band observations which would indicate that the source brightened
by 3.5\,magnitudes in a 1 month period.  While our subsequent optical
images suggest it was either extremely red, $(I-J)\gs 6.5$ or it had
faded by $\gs 4$ magnitudes within 3 weeks of the 2.1-m run.  There are
two possible explanations for the properties of this source: it is
either moving (e.g.\ an asteroid or comet) or highly variable (e.g.\ a
supernova, gamma-ray burst or active galactic nucleus).  We can easily
discount at least one of the variability options -- if the source was a
supernova then it should still be visible in our optical imaging taken
3 weeks after the 2.1-m run.  Equally, by subdividing our 2.1-m imaging
into temporally independent sections and comparing the position of the
object between these, we find no measurable movement of the source,
$\ls 0.5''$, or strong variability over a 6\,hr timespan.  If the
object was moving at the maximum rate allowed by this limit then it
should be still visible within our $z$-band frame taken $\sim 3$ weeks
later.  Its absence therefore rules it out as a solar-system object.
We conclude that \#2549 is most likely a highly variable source,
although unlike the highly-variable AGN found by Gal-Yam et al.\ (2002)
there is no optical or infrared host galaxy underlying \#2549.  We also
note that \#2549 lies 30$''$ from the variable X-ray source in this
field cataloged by Schindler et al.\ (1998) and so is unlikely to be
the same object.

Based on these comparisons we conclude that our
photometrically-selected ERO sample is reliable.

\subsection{The Radio-detected EROs}

We now search for radio counterparts to the EROs in our field.  We
adopt a maximum matching radius of 2$''$ (3 pixels on the
near-infrared images) and search for radio emission above the
3-$\sigma$ limit of 12.6$\mu$Jy centered on the $K$-band centroid of
each ERO.  Of the 68 photometrically-selected EROs, 21 are detected in
our deep VLA observations -- giving a surface density of
$(0.31\pm0.07)$ arcmin$^{-2}$ for EROs with radio fluxes above
12.6\,$\mu$Jy and brighter than $K=20.5$.  The median $K$-band
magnitude of these 21 EROs is $K=19.66\pm 0.22$ (Table~2), almost
identical to the median brightness of the radio-undetected EROs:
$K=19.65\pm 0.09$.  The brightest ERO in the radio has a flux of only
$200\mu$Jy, we can therefore limit the proportion of radio luminous
AGN in the general ERO population to $\ls 2$\% (c.f.\ Willott et al.\
2001).

The radio-detected EROs comprise $31\pm 7$\% of the whole ERO
population down to $K=20.5$ (see also Yan 2001; Mohan et al.\ 2002),
where the error-bar is based on Poisson statistics and hence probably
underestimates the field to field variance in this fraction.  Taking
the expected blank-field radio counts sources down to 12.6\,$\mu$Jy
($\sim 250$, Richards 1999) the ERO counterparts with $K\leq 20.5$
represent $\sim 8$\% of the total radio population.

We plot the cumulative number density of the radio-detected EROs in
Fig.~2. We find, at most, marginal evidence for a difference in the
count slope between the whole ERO population (and given our small
sample also no strong evidence for a break in the count slope with
magnitude) and the radio-detected and undetected samples:
parameterizing the counts as $\log_{10}(N_{\rm ERO}) \propto \alpha K$
we estimate $\alpha=0.59\pm0.05$ for the whole catalog, with
$\alpha=0.51\pm0.08$ and $\alpha=0.64\pm 0.07$ for the radio-detected
and undetected samples.

The radio flux distribution of the radio-detected EROs is
characterized by a strong rise at faint fluxes -- the cumulative
counts being well-fit by a $S_{1.4}^{-0.79\pm0.09}$ power-law.  Hence
the number of EROs detected at 1.4\,GHz rapidly increases at fainter
flux limits: doubling as the flux limit is reduced from $\sim 40\mu$Jy
(typical of previous surveys, e.g.\ Richards et al.\ 1998) to the
12.6-$\mu$Jy limit of the current work. At a flux level of around
50\,$\mu$Jy the ERO counts appears to increase more rapidly (perhaps
indicating the appearance of a new population of very red and very
faint radio sources), but then flattens out at faint fluxes.  We
believe that this flattening is probably a result of the depth of our
available $K$-band imaging which is restricting the identification of
the faintest and typically reddest sources (see Fig.~3).  We see
fairly constant $K$-band to radio flux ratios as a function of radio
flux (see Fig.~3), this is in contrast to the strong decrease in this
ratio with increasing radio flux above 1\,mJy (Waddington et al.\
2000), suggesting a different physical process (star formation, rather
than AGN activity) may be responsible for the faint emission from the
sources in our catalog.

The distribution of EROs in Fig.~3 shows that these galaxies have
typically lower near-infrared to radio flux ratios than average radio
source in the field (many of which are cluster members, Morrison et
al.\ 2002).  The similarity of the K-corrections in these two
wavebands over the redshift range we cover (see \S3.5) suggests that
this is an intrinsic feature of these galaxies.  Assuming that the
local far-infrared to radio correlation holds for these galaxies, then
we conclude that the radio-detected EROs have far-infrared/optical
ratios of $L_{FIR}/L_V\sim 20$--200 (see \S4.2). Locally such high
$L_{FIR}/L_V$ ratios are usually only seen in the advanced stages of
mergers of massive galaxies (e.g.\ Mirabel \& Sanders 1989).  In
comparison, using the median $K$-band magnitude for the
radio-undetected EROs we obtain a limit of $S_K/S_{1.4}\geq 0.6$,
equivalent to $L_{FIR}/L_V\ls 25$ (for an $L_V^\ast$ star-forming
galaxy at $z\gs 1$ the equivalent FIR luminosity would translate into
a limit on the SFR of $\geq 0.1$\,M$_\odot$ stars of $\ls
100$\,M$_\odot$\,yr$^{-1}$).

%
%
\smallskip
\begin{inlinefigure}
\centerline{\psfig{file=f3.ps,width=3.3in}}
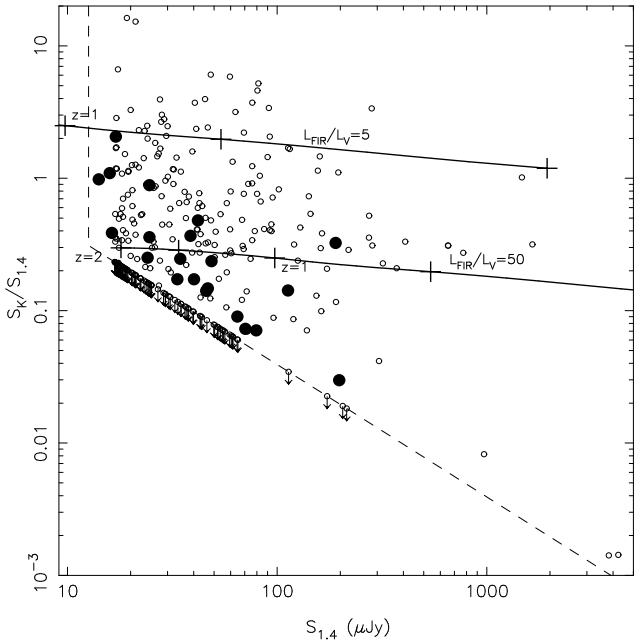
\caption{\small The $S_{K}/S_{1.4\rm GHz}$--$S_{\rm 1.4 GHz}$ plane
illustrating the variation in near-IR to radio flux with radio flux
for all the radio-detected sources in our field. The dashed line shows
the selection boundary for sources brighter than $K=20.5$ and
12.6\,$\mu$Jy and the bold symbols identify those radio sources which
have ERO counterparts.  We also illustrate the tracks for SEDs with
far-infrared to optical flux ratios of $L_{FIR}/L_V\sim 5$ and 50
(using the far-infrared/radio correlation for local galaxies, Condon
1992).  We mark on these tracks the equivalent redshift limits (we
mark $\delta z=0.5$ increments on each track) assuming a model where
the source is an $L_V^\ast (M_V=-21.8)$ galaxy with an SED similar to
a present-day Sbc galaxy and radio luminosities of $L_{1.4}=2.3\times
10^{23}$ (upper track) and $2.3\times 10^{24}$\,W\,Hz$^{-1}$ (lower)
with a radio spectrum of the form $S \propto \nu^{-0.7}$.  These radio
powers are equivalent to SFRs of $\sim 20$ and $\sim
200$\,M$_\odot$\,yr$^{-1}$ for stars above 5\,M$_\odot$, respectively,
or equivalently $10^2$ and $10^3$\,M$_\odot$\,yr$^{-1}$ assuming a
Salpeter IMF which extends down to 0.1\,M$_\odot$ (see \S4.2).  }
\end{inlinefigure}

%
%
\smallskip
\begin{figure*}
\centerline{\psfig{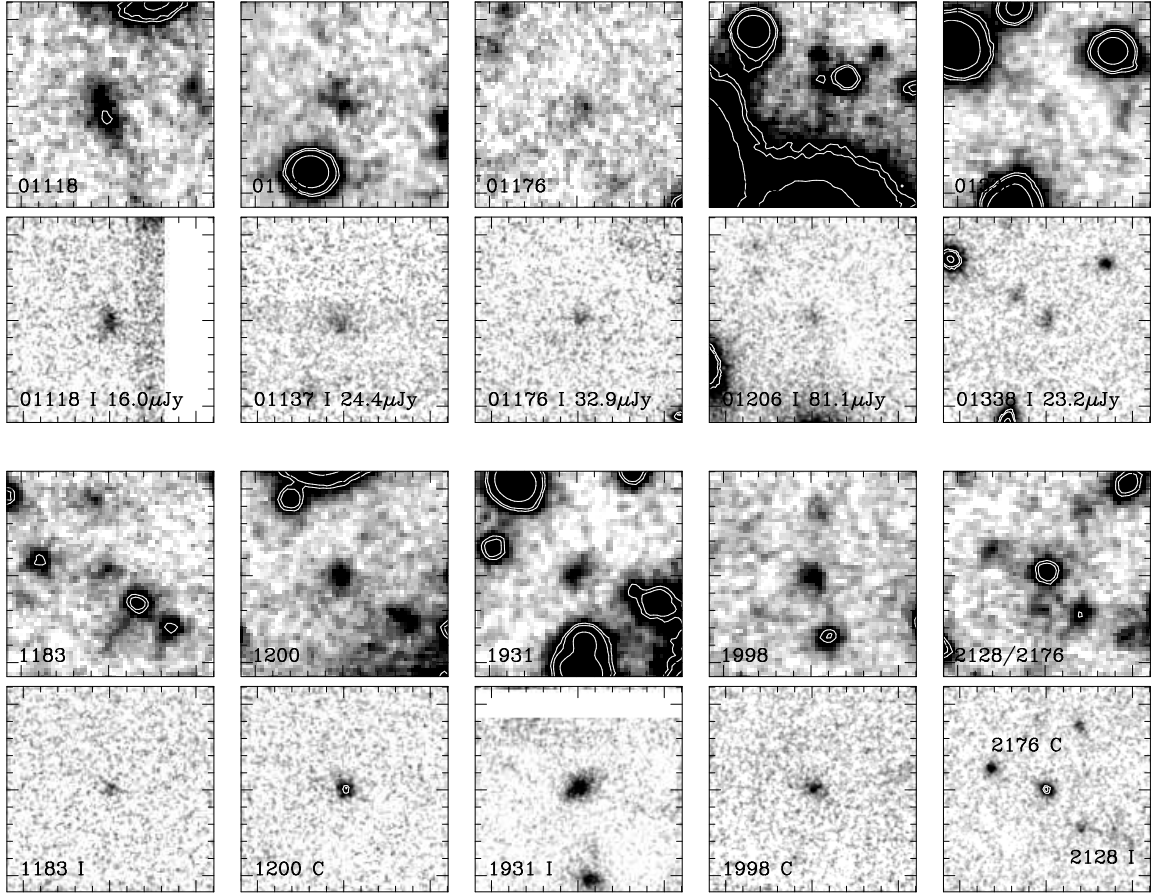}}
\caption{\small Example of the $R$- and $K$-band morphologies of a
random selection of radio-detected (top two rows) and undetected
(bottom two rows) EROs which fall within the UKIRT mosaic. Each field
is shown in two panels, with the Subaru $R$-band images above the UKIRT
$K$-band in both cases. We label the panels with the catalog number,
$K$-band morphology and radio flux (if detected).  There is a clear
tendency for radio-detected EROs to show more complex morphologies,
while radio undetected EROs are more regular and compact. We show two
sources in the panel containing the compact ERO \#2176 and an example
of an amorphous, radio-undetected ERO: \#2128.  Note that \#1931 is the
edge-on disk galaxy identified in the {\it NICMOS} image from Smail et
al.\ (1999b) and \#01208 is the ERO counterpart to SMM\,J09429+4658
(Smail et al.\ 1999a). Each panel is 12$''\times$12$''$ square and has
North top and East to the left, the typical seeing is 0.5$''$ in the
$K$-band and 1.0$''$ in the $R$-band. }
\end{figure*}

Finally, we searched the archival 49.4-ks {\it XMM-Newton}/{\it EPIC}
X-ray image of this field to identify any ERO which may be associated
with a luminous X-ray sources (Alexander et al.\ 2002; Stevens et al.\
2002; Page et al.\ 2002).  We find no such X-ray bright EROs within our
72.7 sq.\ arcmin search region and place typical 3-$\sigma$ limits on
the unabsorbed X-ray emission from individual sources of $<4\times
10^{-16}$\,ergs\,s$^{-1}$\,cm$^{-2}$ in the 2--10\,KeV band.

\subsection{ERO companions}

To search for differences in the local environment of the 
radio-detected and undetected ERO populations we look at the typical
nearest-neighbor distance in our $K<20.5$ photometrically-selected
catalog (Georgakakis et al.\ 1999; Gallimore \& Keel 1993).  We note
that the modest spatial resolution of the KPNO $K$-band images means
this test is relatively insensitive to the presence of close
companions, $<1$--2$''$ seperation.  We find that the median
nearest-neighbor distance is 10.9$''\pm $1.6$''$ for radio-detected
ERO, compared to 15.2$''\pm$0.8$''$ for those EROs which are not
detected in our radio map.  A Kolmogorov-Smirnov test shows only a
10\% chance that the two nearest-neighbor distributions are drawn from
the same parent population.  This suggests a higher frequency of
neighbors for those EROs detected in our radio map, with seperations
roughly corresponding to $\sim 100$\,kpc at $z\gs 1$.  However, these
seperations are probably too large for the companion to be responsible for
dynamically triggering the activity in the radio source.  To
investigate the sub-10-kpc scales, where the signatures of on-going
mergers and strong dynamical interactions will be visible, we need to
look at the detailed morphologies of the EROs (\S3.4).

To further test the association of these neighboring galaxies with the
EROs we analyse their colors. We find that the radio-detected sample
has a median nearest-neighbor color of $(R-K)=3.48\pm 0.20$, while the
closest $K<20.5$ galaxy to the radio-undetected EROs have
$(R-K)=3.96\pm 0.31$ -- in both cases the nearest neighbors have a
median magnitude of $K\sim 19.2\pm 0.2$.  Both of these colors are
marginally redder than the average for a magnitude-matched subsample
from the field: $(R-K)=3.26\pm 0.05$, suggestion that the nearest
neighbors to the EROs may lie at a higher mean redshift, are more
evolved or more obscured than the typical field galaxy.  Both samples
also have comparable fractions of EROs as nearest neighbors -- 14\% of
the radio undetected sample have a nearest neighbor which is also an
ERO (with $(R-K)\geq 5.3$) compared to 12\% for the radio-detected
EROs.  Given that EROs make up only 6\% of the whole $K$-band sample in
this field it is clear that the EROs as a population must be clustered
(Daddi et al.\ 2001; McCarthy et al.\ 2001; Chapman, McCarthy \&
Persson 2000).

\subsection{ERO morphologies}

We can also exploit our high-resolution UKIRT $K$-band and Subaru
$R$-band imaging to quantify the morphological mix within the two
samples of EROs (Fig.~4).  The UKIRT imaging provides restframe
optical morphologies for 19 of the radio-undetected EROs (including
the transient source \#2549, \S3.1, and the star \#2418, \S3.5) and 11
radio-detected EROs.  The properties of these subsamples are
consistent with the full samples (Table~2) with median $K$-band
magnitudes and $(J-K)$ colors of $19.61\pm 0.16$, $2.09\pm 0.12$ and
$19.90\pm 0.24$, $2.23\pm 0.12$ respectively.  In contrast the Subaru
$R$-band image covers the full ERO sample at slightly lower
resolution, but only gives us a restframe UV view of the morphologies
of these galaxies -- which is very sensitive to the presence of both
dust and unobscured star formation.

We visually classify the EROs in both passbands using a simple scheme
designed to crudely distinguish between regular, centrally-concentrated
galaxies, whose extreme colors may arise from a old stellar population,
and more morphologically complex, disturbed or asymmetric galaxies,
where their very red colors could indicate the presence of dust
obscuration.  Following Smith et al.\ (2002a) we define three broad
classes: compact/relaxed systems (``C''); obvious disks, asymmetric or
disturbed/amorphous morphologies (``I'') or too faint to reliably
classify (``F''), see Table~1.  Based on the near-infrared
classifications we estimate that the morphologies of the radio
undetected ERO sample are: 5 compact, 5 disks or disturbed/amorphous
and 7 which are too faint to classify (29\%:29\%:42\%), excluding
\#2549 and \#2418.  For the radio-detected sample the equivalent
numbers are: 0, 9 and 2 respectively (0\%:82\%:18\%).  In both cases
the uncertainties in the numbers in each class are probably 1--2
objects.  We confirm that these two distributions are inconsistent with
being drawn from the same parent population at the $10^{-4}$ level.

The morphological mix of the whole sample is: 18\%:50\%:32\%, identical
to that measured for a $K\ls 21.5$ lensed sample by
Smith et al.\ (2002a).  We find a similar distribution when we use the
restframe UV classifications from our $R$-band images: with the mix for
the full sample being 11\%:52\%:37\%, and the radio-detected EROs
having a slightly higher incidence of disturbed/amorphous morphologies
and a lower incidence of compact images than the radio-undetected subsample.
Thus our sample appears to exhibit a lower fraction of compact, regular
EROs compared to that presented by Moriondo et al.\ (2000), where  80\% of
their EROs were classified as compact/elliptical-like from an analysis of a
somewhat brighter (median $K=19.3$) archival {\it HST} sample.  This
difference could either reflect a mismatch in the classification
schemes; a tendency for the archival {\it HST} fields
to target high-density regions, which may contain a higher fraction
of bulge-dominated EROs; or a real change in the morphological
mix of the ERO population with apparent magnitude (see \S4.2).

Turning to the internal differences within the subsamples of EROs in
our survey, although the samples are small and the classifications are
necessarily crude and somewhat uncertain, it appears that the
radio-detected EROs show a higher incidence of disky and
disturbed/amorphous morphologies compared to the radio-undetected
sample, which in contrast has a higher fraction of compact sources (no
examples of which are seen in the radio-detected sample).  We
illustrate a few examples of the various class of ERO morphologies in
Fig.~4.  We also briefly note the median $(J-K)$ colors of the
different morphological classes (both radio-detected and undetected):
compact, $(J-K)=2.02\pm 0.36$; disky/amorphous, $(J-K)=2.32\pm 0.15$;
faint, $(J-K)=2.09\pm 0.11$.  Thus there is a hint that the compact
population is bluer than the extended (disky, disturbed and amorphous)
subsample, in agreement with the findings of Mannucci et al.\ (2002),
but the large scatter in the colors of these subsamples suggests that
they do not represent homogeneous populations.

Finally, we note that the radio emission from 7 of the radio-detected
EROs, 33\% of the sample, is resolved in our map on scales of
0.8--3.4$''$.  The median intrinsic size of these sources is $1.4\pm
0.4''$, corresponding to $12\pm 3$\,kpc at $z\sim 1$ in our adopted
cosmology.  The remaining 14 radio-detected EROs are unresolved with a
typical limit on the size of their radio emission of $\ls 0.8''$
($<7$\,kpc at $z\sim 1$).  For the 7 resolved EROs we can crudely
compare their morphologies in the radio and optical passbands by
determining the median offset in the position angles of the emission
in these two bands.  We find that the radio and $R$-band emission have
very similar scale sizes and that there is a median offset of only
$11\pm 14$\,degrees between the position angles of the major axes in
the two bands.  This suggests that there is a close relationship
between the extended radio and optical emission in these galaxies and
we return to this point in \S4.

\subsection{ERO Colors}

We begin our investigation of the colors of the EROs in this field by
using the most basic diagnostic diagram proposed for this population:
the $(R-K)$--$(J-K)$ plane.  Pozzetti \& Mannucci (2000) proposed that
evolved and dusty EROs could be separated using this plane -- with the
passive, evolved EROs typically showing bluer $(J-K)$ colors (see also
Smith et al.\ 2002a; Mannucci et al.\ 2002).  We therefore plot the
distribution of all the well-detected EROs (see below for the
definition of well-detected) on this plane in Fig.~5.  Focusing on the
relative distribution of the radio-detected and undetected EROs we see
that the radio-detected EROs tend to fall red-ward, or close to,
the division proposed by Pozzetti \& Mannucci (2000).  In contrast,
the radio-undetected EROs show a much broader distribution, with
roughly equal numbers of galaxies either side of the division (see
also Mannucci et al.\ 2002).  Figure~5 also allows us to classify
\#2418 as a probable star (this is confirmed by the form of its SED in
Fig.~6).

To investigate the photometric properties of the ERO populations in
more detail we now exploit our multiband photometry.  We begin by
simply constructing average colors for each sample.  The median colors
of the radio-detected and radio-undetected ERO samples are given in
Table~2 (where the errors are the uncertainty on the median from
bootstrap resampling).  We find that the radio-undetected EROs are
typically bluer than the radio-detected sample in colors long-ward of
$\sim 1\mu$m -- consistent with the color distribution of the two
samples in Fig.~5. We also find that the radio ERO population may be
slightly more homogeneous than either the whole ERO catalog or the
radio-undetected sample.  For example, the radio-undetected sample
exhibits a broader spread of colors with a  typical dispersion 
of $0.44\pm 0.03$ around the median color (Table~2), compared
to $0.37\pm 0.03$ for the radio-detected sample (this is also
illustrated by the broader range of SEDs for the radio-undetected sample in
Fig.~6). The varied mix of colors for the radio-undetected EROs is not
surprising as we expect that this sample probably contains both
passive evolved galaxies and dusty active systems whose radio flux is
below the sensitivity limit of our VLA map.  The presence of passive,
evolved galaxies in this sample is further supported by the more
abrupt change in the colors of this sample at $\ls 1\mu$m,
corresponding to the 4000\AA\ break at $z\ls 1.5$.  However, the
relatively modest differences between the median colors of the radio
detected and undetected samples also suggests that the
radio-undetected ERO sample contains a reasonable fraction of galaxies
with dusty SEDs (or that the relative redshift distributions of the
two samples conspire so that the observed colors are more similar than
they are in the restframe).

%
%
\setcounter{table}{1}
\begin{table*}
{\footnotesize
\begin{center}
\caption{\hfil Median Colors of ERO Subsamples \hfil}
\begin{tabular}{lccccccccl}
\hline\hline
Sample & $N$ & $K$& $(R-K)$ & $(I-K)$ & $(z-K)$ & $(J-K)$ & $(H-K)$ & Dispersion & Note \cr
\hline
\noalign{\smallskip}
\multispan3{Radio~detected \hfil }\cr
All & 21 & $19.66\pm 0.22$ & $5.88\pm 0.13$ & $4.53\pm 0.15$ & $3.56\pm 0.08$ & $2.05\pm 0.09$ & $1.18\pm 0.08$ & $0.37\pm 0.03$ & \cr
\noalign{\smallskip}
\multispan3{Radio~undetected \hfil }\cr
All & 45 & $19.65\pm 0.09$ & $5.82\pm 0.11$ & $4.67\pm 0.14$ & $3.40\pm 0.07$ & $1.98\pm 0.06$ & $1.05\pm 0.10$ & $0.44\pm 0.03$ & Excluding \#2418 \& \#2549  \cr
\noalign{\smallskip}
Dusty SED    & ~9 & $19.67\pm 0.10$ & $5.73\pm 0.19$ & $4.51\pm 0.13$ & $3.39\pm 0.18$ & $2.53\pm 0.17$ & $1.21\pm 0.11$ & $0.36\pm 0.03$ & \cr
Evolved SED  & 11 & $19.60\pm 0.11$ & $5.59\pm 0.17$ & $4.89\pm 0.20$ & $3.49\pm 0.11$ & $1.97\pm 0.14$ & $1.18\pm 0.12$ & $0.41\pm 0.07$ & \cr
Badly fit    & ~8 & $19.22\pm 0.24$ & $5.73\pm 0.32$ & $4.27\pm 0.22$ & $3.60\pm 0.37$ & $1.81\pm 0.12$ & $0.83\pm 0.04$ & $0.49\pm 0.15$ & Excluding \#2418 \& \#2549 \cr
\noalign{\smallskip}
\multispan3{Radio~detected \& undetected \hfil }\cr
Not fit      & 20 & $19.92\pm 0.12$ & $5.81\pm 0.13$ & $4.94\pm 0.17$ & $4.48\pm 0.45$ & $2.19\pm 0.19$ & $1.15\pm 0.15$ & ... & \cr 
\hline
\end{tabular}
\end{center}
}
\end{table*}

\subsubsection{Photometric classification of EROs}

To further investigate the photometric classifications of individual ERO
we have undertaken a simple test to determine whether the colors of a
particular ERO are better described by a relatively feature-less,
dust-dominated SED or a spectrum with a strong break.  The simplest
way to apply this test is to employ the {\sc hyperz} photometric
redshift code (Bolzonella et al.\ 2000) to compare the ERO colors to
two different families of SEDs: dusty, young starbursts (or AGN) and
almost dust-free evolved systems.  For the dusty AGN we assume that
their optical and near-infrared colors are dominated by a co-eval
starburst (which produces the dust) -- although the AGN may contribute
significantly to the radio emission from the system.  By relying on
just two simple classes of SED we will clearly fail to describe any
systems with more complex mixes of evolved stars and obscured and
unobscured young stars (a situation which may be common place in many
of these systems at high-$z$, e.g.\ the edge-on bulge-strong disk
galaxy ERO \#1931, Fig.~4 and Table~1).  However, by concentrating on
those EROs which are well-fitted by these simple models we can
determine the relative proportion of EROs at the extremes of these
classes.

For this test we employ a single star formation model described by an
exponential declining star formation rate with a e-folding time
($\tau$) of 1\,Gyr, a Miller-Scalo IMF and solar metallicity. This
model can provide an adequate description of the colors of luminous
elliptical galaxies at the present-day (Bolzonella et al.\ 2000) and
hence is a reasonable choice given the expectation that both the
passive and star-forming ERO populations may represent the precursors
of the elliptical galaxies seen in the local Universe.  As a test of the
sensitivity of our classifications to the details of the star
formation model we have also used an instantaneous burst and a
$\tau=2$\,Gyr model in place of the $\tau=1$\,Gyr model and find that
the proportions of EROs in the different classes varies by less than
10\%.

In our model the initial formation redshift, the dust reddening and
the redshift of the ERO are all free parameters subject to the
following constraints.  For the starburst/AGNs the reddening, based on
a Calzetti et al.\ (2000) dust model, can vary in the range
$A_V=1$--6, consistent with the range of reddening estimated for their
spectroscopically-identified star forming EROs by Cimatti et al.\
(2002). In contrast for the passive models it is required to be
$A_V<0.5$ -- hence by restricting the range of possible reddening we
isolate the two families of solutions which are capable of fitting the
very red colors of the EROs: either unobscured but old stellar
populations or highly obscured, actively star forming systems.  In
both cases the redshift of the ERO is allowed to vary between
$z=0$--3, a large enough range to comfortably encompass all of the
EROs in Cimatti et al.\ (2002).  The upper limit on the redshift
removes the possibility of fitting infeasibly high redshifts due to
the 4000\AA\ break being confused with the Lyman break, although we
note that at least one ERO in our sample has a proposed redshift of
$z>3.4$ (\#01206, Smail et al.\ 2002).

To obtain useful constraints from our analysis, we first require that
the ERO is detected in at least 3 passbands at more than 3-$\sigma$,
this reduces the samples to 17 radio-detected EROs and 30 radio
undetected EROs.  We list at the bottom of Table~2 the median $K$-band
magnitude and colors of those EROs which were removed from our analysis
by this signal-to-noise ratio (SNR) cut.  As expected these are fainter
on average, $K=19.92\pm 0.12$, than the typical ERO in our sample,
which has $K=19.63\pm 0.09$.  This magnitude difference is equivalent
to that expected for a change in the median distance modulus of the
population of $\delta z\sim 0.1$. As many of these EROs are undetected
in a particular passband we have improved the signal-to-noise of their
colors by calculating these from the median fluxes in the various
passbands, including cases where the calculated flux is negative
(Table~2).  We find that these colors are consistent with those for the
full samples, with the largest discrepancy being the $(z-K)$ color
which differs by $\sim 2.2\sigma$.  We conclude that the EROs removed
by our SNR cut are at most slightly redder than the sample we retain,
and the main reason for their lower SNR is that they are simply fainter
on average.

We fit the $RIzJHK$ photometry (including limits) of the high-SNR
radio-detected and radio-undetected samples to both families of model
SEDs allowing for $\delta m=0.05$ systematic errors on the
calibrations of the different passbands.  We then compare the relative
$\chi^2$ of the best-fit models in both cases and classify the EROs on
this basis into either evolved, ``E'', dusty, ``D'', or badly-fit,
``B'' (see Table~2).  We plot the SEDs of the sources used in the
analysis in Fig.~6. Encouragingly this test identifies \#2418 and
\#2549 as pathological SEDs which are not well fit by either galaxy
model ($\chi^2\gg 100$) and we remove these from our sample,
reducing it to 28 sources (Table~2).  

Restricting our comparison to those EROs which are reasonably well-fit
by either SED, $\chi^2<2.7$ (90\% confidence limit) we find that 12
from the 17 EROs in the high-SNR, radio-detected sample are
well-fitted, 11 of these by the dusty SED (65\%).  This supports the
suggestion that the bulk of this population represent dusty, active
galaxies and gives us good confidence that this model SED can be reliably
used to identify dusty systems in the high-SNR, radio-undetected
sample.  We find only one radio-detected ERO which is better fit by an
evolved SED, this is \#11077, a relatively faint radio source with an
irregular ERO counterpart. Examining the SED of this source we see
that it is unique amongst the radio-detected EROs in having a blue
$(R-I)$ color, which cannot be well-fitted by a dusty SED (Fig.~6).
Comparing the morphology of the ERO in the $R$- and $K$-bands we find
that the galaxy comprises two regions -- a blue area to the east of
the radio source and a very red area to the west -- this complex
morphology suggests that the simple model we use to fit may not be
applicable to this galaxy.

The five high-SNR, radio-detected EROs which are not well-fit by the
dusty SED (\#01045, \#01105, \#01137, \#11018 and \#12861) are ill-fit
by either family of model SEDs ($\chi^2>5$--11).  This could result
from systematic errors in our photometry for these galaxies (although
we have checked that their colors do not appear to be strongly
effected by nearby galaxies), or because their SEDs are not
well-matched to the two simple alternatives used in our analysis:
e.g.\ composite systems with evolved stellar populations and AGN
contributions, galaxies with complex dust geometries, or very high-$z$
starbursts which are also not included in our model (e.g.\ Waddington
et al.\ 1999).  We will discuss the properties of these sources in the
context of a fuller photometric analysis of the complete radio
population in this field in Owen et al.\ (2002).

Turning to the radio-undetected EROs we find that 20 of the 28 sources
(79\%) are reasonably described by one or other of the two families of
model SEDs.  Of these, 11 (39\%) are better fit with a passive,
evolved SED, compared to 9 (32\%) which are well-fitted by a dusty
SED.  The median $K$-band magnitudes of the two classes are
$K=19.60\pm 0.11$ for the evolved subsample and $K=19.67\pm 0.10$ for
the dusty sources, if we include with the latter the 11 well-fit 
dusty, radio-detected EROs ($K=19.91\pm 0.28$) we obtain a median magnitude of
$K=19.82\pm 0.09$ for all of the EROs whose colors are well-fit by a
dusty SED. Thus there is a slight tendency for the EROs with dusty
SEDs to be fainter than those with evolved SEDs.  The 8 radio
undetected EROs which are not well-fit with either family of model
SEDs (ignoring \#2418 and \#2549) are surprisingly bright on average
$K=19.22\pm 0.24$. To test whether this results from an error in the
relative calibration of our photometry we increase the estimate of the
systematic errors in the relative calibration of the photometry from
$\Delta m=0.05$ to 0.2 and confirm that this does not
significantly effect the average magnitude of the badly-fitting SEDs,
although it does reduce their number from 8 to 4 -- with the EROs
making up the difference being equally classed as dusty and evolved
systems.  We also confirm that photometry of these badly-fit EROs is
not unduly affected by crowding.

%
%
\smallskip
\begin{inlinefigure}
\centerline{\psfig{file=f5.ps,width=3.3in}}
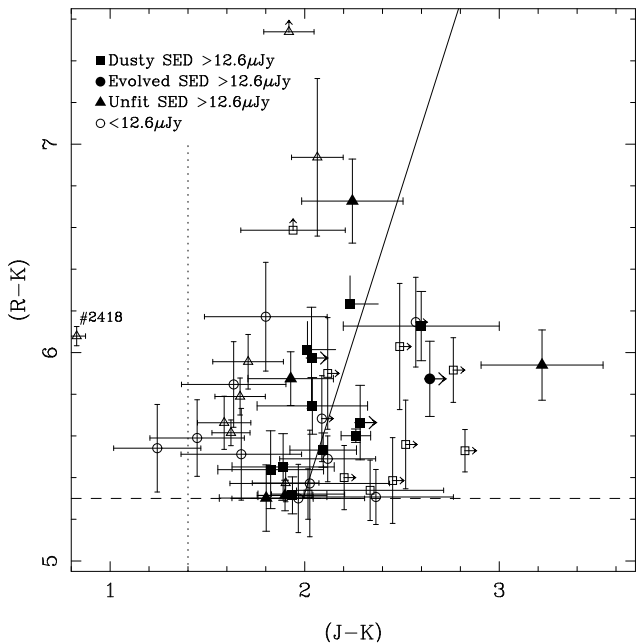
\caption{\small $(R-K)$--$(J-K)$ color-color diagram for the 47
high-SNR ERO's brighter than $K=20.5$ used in the SED fitting. We
identify radio-detected and undetected EROs using filled and open
symbols and in addition code the points depending upon the better
fitting SED: either dusty, evolved or badly-fit. The solid line
represents the boundary between dusty starburst and evolved, passive
EROs proposed by Pozzetti \& Mannucci (2000) -- the dusty galaxies
should lie to the right of the line (redder $(J-K)$ colors), with the
evolved systems on the left. Note that the high-SNR, radio-selected
EROs lie either close to, or to the right, of this division, as do the
majority of the high-SNR, radio-undetected EROs with dusty SEDs,
confirming that the proposed boundary is reasonably effective. The
dotted line divides low-mass stars from galaxies (which are redder in
$(J-K)$). The dashed line shows the $(R-K)=5.3$ boundary used to
define our ERO sample.  }
\end{inlinefigure}

%
%
\begin{figure*}[tbh]
\centerline{\psfig{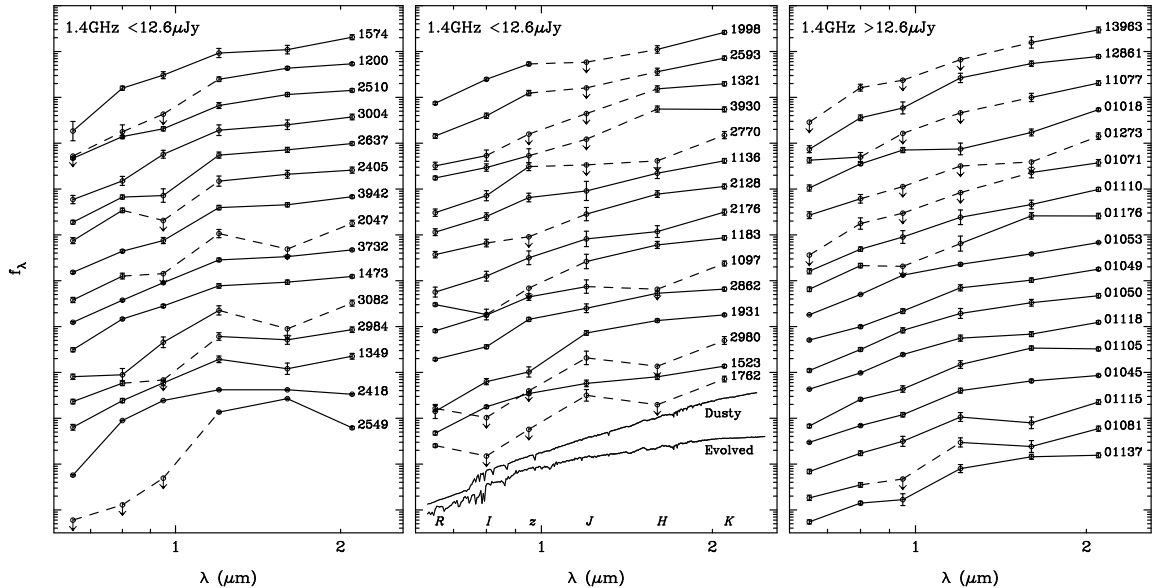}}
\caption{\small The broad-band SEDs for the high-SNR, radio-detected
and undetected EROs from the sample used in our photometric analysis,
each panel is ordered in terms of increasing $(J-K)$ color. The
left-hand and middle panels show EROs which are undetected in our radio
map, while the right-hand panel shows the SEDs for the 17 high-SNR,
radio-detected EROs.  These SEDs are plotted on an arbitrary flux scale
and we indicate uncertain regions of the SEDs by dashed lines.  We also
show an example of a dusty starburst SED and a passive, evolved galaxy,
both at $z=1$ at the bottom of the central panel, as well as the
central wavelengths of the 6 passbands used in our analysis. Note the
peculiar SEDs of ERO \#2418 and \#2549, we identify \#2418 as a
late-type M dwarf (as suggested by its colors in Fig.~6), while the
comparison of our independent $J$- and $K$-band observations of the
field shows that \#2549 is a transient source.}
\end{figure*}

The median colors of the various subsamples of radio-undetected EROs
are given in Table~2.  The dispersion of the colors of the individual
EROs in each subsample around their median are also listed there, these
show that the dispersion within the subsamples of radio undetected EROs
classed as dusty or evolved is less than that seen for the whole
sample.  This confirms that as expected the photometric classifications
isolate more homogeneous subsamples from the overall radio-undetected
ERO population, with the exception of the badly-fit subsample whose
larger dispersion suggests that it still contains a mix of SED types.

\subsubsection{The mix of dusty and evolved EROs}

The strong preference for fitting the colors of high-SNR
radio-detected EROs with the dusty SED model, 11/12 well-fit examples,
gives us confidence that this model can be reliably employed to
identify similarly dusty, active systems in the radio-undetected ERO
sample and hence estimate the prevalence of such galaxies in the full
ERO population.

Taking the estimate that at least 20\% (9/45) of the radio-undetected
EROs have dusty SEDs and including the 11 high-SNR, radio-detected
EROs with similar photometric classifications, we conclude that at least
$\geq 44$\% (20/45) of those EROs with well-sampled SEDs have colors
compatible with dusty starbursts, or $\geq 30$\% (20/66) of {\it all}
EROs with $(R-K)>5.3$ and $K<20.5$ (median $K\sim19.6$).  This is
consistent with the estimate of the proportion of dusty starbursts at
$K<19.2$ (median $K\sim 18.3$) of 33--67\% from Cimatti et al.\
(2002).

There are two different assumptions which can now be made to broaden
this constraint: one addresses the failure rate for fitting dusty EROs
with our model SED, the other with the number of dusty systems in the
low-SNR ERO sample.  Firstly, we could assume that the proportion of
dusty SEDs in the high-SNR radio-detected sample is the same as that
measured for those EROs with well-fitted SEDs, 92\% (11/12), i.e.\ that
the majority of the radio-detected EROs have dusty SEDs.  Based on this
we can then estimate the failure rate for fitting the colors of a dusty
ERO with our model template as 5/16.  Adopting the same failure rate
for the dusty galaxies in the high-SNR radio-undetected sample, which
have similar quality photometry, means that half of the badly-fit EROs
are actually dusty galaxies (we would get the same estimate if we had
simply proportioned the badly-fit ERO sample on the basis of the
relative numbers of the successful fits). Hence we would estimate that
at least 46\% (13/28) of the high-SNR radio-undetected EROs have dusty
SEDs.  Adding in the radio-detected EROs then leads to a lower limit of
$\gs 50$\% on the proportion of dusty starbursts in the full ERO
sample.  Secondly, we could assume, based on the overall similarity of
their median colors, that the high-SNR EROs are representative of the
full samples.  If this assumption is adopted then we would again
conclude that a minimum of $\gs 44$\% of the EROs in our full sample
have colors which are well-fit by a dust-reddened starburst.

Hence, either of these assumptions would suggest that a minimum of
about a half of the $(R-K)>5.3$ and $K<20.5$ ERO population have
colors consistent with those expected for dusty starbursts.  As these
two assumptions are independent, we can also combine them by applying
the mis-classification correction to the radio-detected and undetected
samples (giving 16/17 and 13/28) and then assuming that the relative
proportions of EROs of the photometric types in the high-SNR samples
apply independently to the full samples (yielding 20/21 and 21/45).
In this way we estimate that perhaps $60\pm 15$\% of the full ERO
sample are distant, dusty starburst/AGNs, with the remainder being
passive systems (although these may also include a small number of
very high-redshift galaxies, e.g.\ Waddington et al.\ 1999).  This
estimate of the proportion of evolved and dusty EROs in the $K<20.5$
population is of course also sensitive to the strong clustering of the
passive ERO population.

Our photometric classification also provides some other information about
the EROs: in particular the best-fit redshift and reddening.  However,
the limited precision of our photometry and the strong degeneracies
between these parameters means that they should be viewed with
caution.  For the high-SNR, radio-detected EROs, which are well-fit by
the dusty SED, the median redshift is $z=1.0\pm 0.3$, with a quartile
range of $z=0.8$--1.5.  This range is very similar to that seen in the
spectroscopic survey of Cimatti et al.\ (2002) and gives us some
confidence in the average properties we determine for this population.
The median reddening for this sample is $A_V\sim 3.5\pm 0.4$.  For the
radio-undetected EROs, those galaxies which are fitted with a dusty
SED give a similar median redshift of $z=1.0\pm 0.2$ and a slightly
lower reddening of $A_V\sim2.0 \pm 0.5$.  These estimated reddenings
are consistent with that estimated for their composite spectrum of
star-forming EROs by Cimatti et al.\ (2002), $A_V\sim 2.4\pm 0.9$.
For both subsamples the median $K$-band luminosities at the estimated
redshifts corresponds to $M_V\sim -18.4\pm 0.4$, or $M_V\sim -21.5$
when corrected for reddening.  The unobscured luminosities of these
galaxies would be comparable to $L_V^\ast$ galaxies today, however,
their young stellar populations would be expected to fade
substantially and so it is clear that these galaxies are likely to
evolve into sub-$L_V^\ast$ galaxies at the present-day, unless the
current starburst continues for $\gs 10^8$\,yrs.

Next, we look at the relationship between our crude restframe visual
morphologies and the preferred photometric classification for 28 EROs (8
radio-detected and 20 radio-undetected, Table~1).  The morphological
mix in both cases is similar to the proportions of the different SED
types, if we interpret the compact EROs as passive, evolved systems
(Mannucci et al.\ 2002). For the radio-undetected sample we find a
broad equality between the classes: 29\%:29\%:42\% for
compact:disky/amorphous:faint, similar to 39\%:32\%:29\% for the
evolved, dusty and badly-fit SEDs.  Similarly, for the radio-detected
sample, which is dominated by dusty SEDs, the morphological mix is:
0\%:82\%:18\% compared to the distribution of SED fits: 6\%:65\%:29\%.
However, a one-to-one comparison for individual EROs shows that while
the distributions appear to be telling the same story, the individual
galaxies are much more mixed: half of those EROs with dusty SEDs have
disky, amorphous or disturbed morphologies, but none of the EROs with
passive SEDs have compact morphologies (this morphological class is
almost evenly split between dusty SEDs and failed fits).  We also note
that there is no clear trend in the SEDs of those radio-detected EROs
which are resolved in our VLA map.

To summarize, we place a firm lower limit of $\geq 30$\% on the
proportion of dusty starbursts in the full $(R-K)>5.3$ and $K<20.5$
ERO population.  Adopting two separate corrections for the number of
dusty EROs which are missed either due to the failure of our photometric
classification or because they were removed due to their poor
photometry, we propose that $\gs 45$--50\% is a more realistic lower
limit on the fraction of dusty EROs.  Finally, by combining these
corrections we suggest that the actual fraction of EROs with dusty
starburst/AGN-like colors is probably close to $60\pm 15$\%.

\section{Discussion}

\subsection{The evolution of dusty and evolved EROs}

We have estimated the relative proportion of dusty, active and passive,
evolved systems in the $K\leq 20.5$ ERO population. By comparing this
to the mix of classes estimated for the $K\leq 19.2$ sample by Cimatti
et al.\ (2002) we can identify any decline in either class of EROs
which might be responsible for the apparent break in the ERO counts at
$K\sim 19$--20 (McCarthy et al.\ 2001; Smith et al.\ 2002a).  The
similarity of the two estimates suggests that there is no evidence for
a significant shift in the mix of these two sub-populations across
$K\sim 18$--20, although it should be noted that the Cimatti et al.\
sample is selected using a slightly bluer color criteria, $(R-K)\geq
5$, which complicates this comparison. This apparent lack of a
substantial shift in the nature of the ERO population may indicate that
the break arises instead from the volume sampled -- with $K\ls 20$ EROs
populating a relative narrow redshift range.  This volume would be
bounded at the low redshift end by the requirement that the galaxy is
observed at high enough redshift, i.e.\ sufficiently far into the
restframe UV, to enable a dust-reddened starburst or evolved stellar
population to appear extremely red, $(R-K)\geq 5.3$ (Daddi et al.\
2001; Firth et al.\ 2002).  Observationally this appears to require
$z\gs 0.8$--1 (\S3.5; Cimatti et al.\ 2002).  At the upper-end the
redshift limit may be imposed either by the luminosity function,
passive bluing of the stellar population, or the increasing fraction of
unobscured star formation in both sub-classes of ERO (the continuity
between high-$z$ SCUBA galaxies and the radio-selected ERO population
suggested below would argue against the latter suggestion). If the
break in the counts is due to the surveys reaching beyond the volume
containing the ERO population then the $K\sim 19$ break (Fig.~2) should
roughly correspond to $M^\ast$ for a composite luminosity function with
a Schechter form. This would imply that the characteristic luminosity
of the ERO population is very similar to that seen in the local field,
$M^\ast_V\sim -21.5$, and that the ERO luminosity function has a faint
end slope, $\alpha$, which is shallower than $\alpha \sim -0.7$ (the
slope of the differential counts beyond the break).  To illustrate this
we plot a toy model in Fig.~2, this has a mix of Gaussian luminosity
functions each with mean luminosity of $M_K=24.3$ ($M^\ast+0.7$ at
$z=0$, Cole et al.\ 2001) and a dispersion of $\sigma_K=1$ (similar to
the width of the luminosity function of elliptical galaxies
locally). These are combined to produce a redshift distribution with a
mean redshift of $<\! z\!>\sim 1.1$ and a dispersion of $\sigma_z=0.2$
(Cimatti et al.\ 2002; Firth et al.\ 2001), and the cumulative counts
calculated using a simple K-correction.  Clearly this toy model (with
some further tuning) can reasonably reproduce the ERO counts,
indicating that a combination of a narrow luminosity function and a
narrow redshift range are a sufficient, if not unique, explanation of
the form of the counts of the whole ERO population across $K\sim
17$--21.

However, such a model has two obvious shortcomings -- firstly, it
requires that the evolutionary behavior of the dusty and passive ERO
populations is similar, which seems highly unlikely, and secondly the
required evolution means that the dusty, active EROs selected in our
radio map are distinct from the higher redshift and fainter
counterparts to the SCUBA galaxies, rather than representing a simple
continuity of the population to lower redshifts and brighter $K$-band
fluxes.  For these reasons we reject this model and construct a second
toy model.

For the second model we assume that the number counts of dusty EROs
are well-described by the power-law slope estimated for the radio
detected population in \S3.1, but with a normalization increased to
reproduce the $\sim 60$\% fraction of the $K<20.5$ ERO population which we
estimate have dusty SEDs.  To this we add a Gaussian population
representing the passive sub-population and normalize the combined
sample to reproduce the weighted cumulative surface density of
EROs at $K<20.5$ from Thompson et al.\ (1999) and Smith et al.\
(2002a).  Representative parameters for the passive sub-population
require a mean magnitude of $K=19.1$ ($M_K^\ast+0.5$ at $z=0$ assuming
no evolution, or $M_K^\ast+1.5$ assuming passive evolution from $z\sim
1.1$) and a FWHM of 1.2\,mag.  We illustrate the cumulative counts of
both sub-populations and the combined population in Fig.~2.  We note
that this model predicts that the relative proportions of dusty and
evolved EROs for a $K<19.2$ sample is 35\%:65\%, in agreement with the
current limits from Cimatti et al.\ (2002).  Clearly this model, even
without elaborate tuning, can provide an adequate description of the
current data, while at the same time providing continuity between the
dusty EROs detected in our VLA map and the higher redshift and fainter
counterparts to submm sources (Smail et al.\ 2002; Ivison et al.\
2002) and we therefore prefer it.  A clear distinction between the
alternative models will come from extending the counts of EROs
fainter, to $K\gs 21$--23, and estimating the relative mix of evolved
and dusty systems at these depths.  However, the depth of the imaging
required for such a survey, $R\sim 28$ and $K\sim 23$, will make it
observationally demanding.

\subsection{Star-formation in the dusty ERO population}

The resolution of over a third of the radio-detected EROs in our VLA
map, along with the close correlation between the radio and optical
morphologies of these EROs, suggests a common mechanism for the
emission at these two wavelengths.  The simplest explanation is that the
emission in both bands is powered by massive star formation 
within an extended region -- on a scale of $\sim 10$\,kpc.   Thus for
at least a third of the radio-detected EROs, star formation is likely
to contribute the majority of the radio emission (this includes one of
the brightest radio EROs, \#01053).  Furthermore, we note that using
the median 1.4-GHz surface brightnesses of these resolved EROs
and the star formation calibration discussed below, they are
expected to have high star formation surface densities: 
$\sim 4$M$_\odot$\,yr$^{-1}$\,kpc$^{-2}$, more than sufficient
to drive powerful winds. 

\subsubsection{The fraction of AGN-dominated EROs}

To reliably quantify the star formation occurring in the complete
radio-detected ERO sample we must estimate the contamination from
AGN-dominated systems.  The AGN fraction in the whole $\mu$Jy radio
population can be determined from the spectroscopic and morphological
surveys of $\mu$Jy radio sources (e.g.\ Muxlow et al.\ 1999; Roche,
Lowenthal \& Koo 2002).  These suggest that at most 20\% of radio
sources at $\gs 100\mu$Jy are likely to be AGN.  However, these
estimates are typically based on samples of radio sources with brighter
optical counterparts than the ERO sample analysed here (and the AGN
tend to be the bluer sources in even these samples, Roche, Lowenthal \&
Koo 2002; Waddington et al.\ 2000).  Nevertheless, non-thermal emission
is seen in a handful of well-studied EROs (Pierre et al.\ 2001; Willott
et al.\ 2001; Smith et al.\ 2001; Cowie et al.\ 2001) indicating that
AGN are not completely absent from the ERO population, but the fraction
of these in complete photometrically selected ERO samples is not well
determined (Cimatti et al.\ 2002).  The most useful constraint on this
fraction for our purpose is that estimated from the number of ERO
counterparts to faint X-ray sources (Alexander et al.\ 2001, 2002;
Brusa et al.\ 2002; Mainieri et al.\ 2002).  While we earlier confirmed
that none of the EROs in our sample is associated with a luminous X-ray
source, using archival {\it XMM-Newton} observations, much more
stringent limits are now available on X-ray counterparts to EROs from
recent {\it Chandra} and {\it XMM-Newton} observations of blank fields
(which have lower backgrounds than our cluster field).

Alexander et al.\ (2002) identify six $(I-K)\geq 4$ and $K\leq 20.1$
X-ray sources in a 70.3 sq.\ arcmin region within the 1-Ms {\it
Chandra} Deep Field North exposure, giving a surface density of
$0.085\pm 0.035$ arcmin$^{-2}$.  Based on the relative numbers of
$(I-K)\geq 4$ and $(R-K)\geq 5.3$ sources and the number counts of
$(R-K)\geq 5.3$ EROs we would expect roughly $\sim 0.07\pm 0.03$
arcmin$^{-2}$ X-ray-detected EROs with $(R-K)\geq 5.3$ and $K<20.5$.
This estimate assumes that the X-ray-detected fraction is not a strong
function of color or $K$-magnitude, although we note that 2/6 sources
in the Alexander et al.\ (2002) sample have $(I-K)=4.0$ and hence are
borderline cases.  Compared to our ERO surface density of
$(1.04\pm0.12)$ arcmin$^{-2}$ this indicates that around $\sim 10$\% of the
ERO population are detected in very deep X-ray observations.

However, the {\it Chandra} exposure used by Alexander et al.\ (2002)
is sufficiently sensitive that it will detect not only obscured AGN
sources, $L_X\gs 10^{44}$\,ergs\,s$^{-1}$, but also powerful star
forming galaxies at $z\sim 1$ (Ranalli et al.\ 2002).  If we instead
restrict the X-ray sample to those EROs which are too X-ray luminous
to be ULIRG-like starbursts at $z\sim 1$, i.e.\
$L_X($0.1--2.4\,KeV$)\gg 4\times 10^{42}$\,ergs\,s$^{-1}$ (twice the
X-ray luminosity of the local X-ray-bright starburst NGC\,3256, which
has $\log_{10} (L_{FIR}/L_\odot) \sim 11.8$, Moran et al.\ 1999), then
the number of probable AGN-dominated sources in the Alexander et al.\
(2002) sample would drop to three (two of which have borderline,
$(I-K)=4.0$, colors), or an estimated surface density of $\ls 0.03$
arcmin$^{-2}$ using our ERO criteria.  We conclude that at most 3\% of
the ERO population with $(R-K)\geq 5.3$ and $K<20.5$ are likely to
host energetically dominant AGN.  This is consistent with the
estimates from ERO counterparts to hard X-ray sources detected by {\it
XMM-Newton} in Franceschini et al.\ (2002), as well as the recent
estimate that 1--2\% of the ERO population have hard X-ray
counterparts by Brusa et al.\ (2002) and Mainieri et al.\ (2002).

While the occurrence of AGN-dominated sources in the ERO population is obviously low, the
estimates derived above do not tell us how frequently radio-selected
EROs are detected in the X-ray waveband.  Fortunately, we can also
crudely estimate this from the Alexander et al.\ (2002) survey.  Three
of the X-ray-detected sources with $(I-K)\geq 4$ in Alexander et al.\
(2002) are also detected in deep 1.4-GHz observations, two of these
have X-ray and radio emission consistent with an X-ray bright
starburst at $z\sim 1$, suggesting a surface density of $\sim
0.02$\,arcmin$^{-2}$ for X-ray-detected AGNs with $\mu$Jy ERO hosts,
corresponding to $\sim 6$\% of the whole radio-detected ERO
population.  This rate of AGN contamination is consistent with our
identification of only one radio-detected ERO with an evolved SED
(assuming the radio emission from this galaxy arises from a weak AGN,
see \S3.5), corresponding to 6\% of the sample.  We therefore adopt
6\% as the AGN contamination in our sample and conservatively assume
that the radio emission from any EROs hosting an AGN is dominated by
the AGN.

\subsubsection{The star formation density in EROs}

We can now use our ERO sample to estimate the star formation density
in the most obscured starbursts at $z\sim 1$ (Cram et al.\ 1998;
Oliver et al.\ 1998; Haarsma et al.\ 2000).  We calculate two
estimates of the star formation density -- first we take the minimum
number of galaxies in our sample which are likely to be dusty,
active systems based on the analysis of their optical-near-infrared
SEDs: this subsample comprises 11 high-SNR, radio-detected EROs and 9
radio non-detections, and provides a conservative lower bound on the
estimated star formation density.  

Those radio-undetected EROs with dusty SEDs are likely to either be
more distant or have typically lower star formation rates (and hence
radio fluxes) than the radio-detected examples.  Based on the analysis
in \S3.5.2 there is a hint that these galaxies have similar redshifts
to the radio-detected EROs, but lower reddening's -- and hence perhaps
lower star formation rates (Hopkins et al.\ 2001).  Using their
estimated surface density and extrapolating the cumulative number
counts of radio-detected EROs in our field from \S3.2, we would expect
most of the dusty, radio-undetected EROs to have radio fluxes above
$\sim 5\mu$Jy and hence we assume that the typical dusty,
radio-undetected ERO has a radio flux of $\sim 10$\,$\mu$Jy.

For the second estimate we adopt the incompleteness estimates for the
photometric classification from \S3.5 and hence use 20 radio detections
and 21 radio-undetected EROs.  Simply summing the 1.4-GHz fluxes for
these two subsamples we obtain a total radio flux of $\geq 910$\,$\mu$Jy
for the conservative case and 1370\,$\mu$Jy for the more realistic
estimate (these include contributions from the radio-undetected
starbursts of 90\,$\mu$Jy and 210\,$\mu$Jy respectively).  Applying
our estimated AGN contamination reduces both of these estimates by
6\%, however, if this contamination preferentially effects the most
luminous, unresolved radio EROs then the radio fluxes could be reduced
by as much as $\sim 15$\% (roughly half the difference between our
``conservative'' and ``realistic'' cases).

%
%
\smallskip
\begin{inlinefigure}
\centerline{\psfig{file=f7.ps,width=3.3in}}
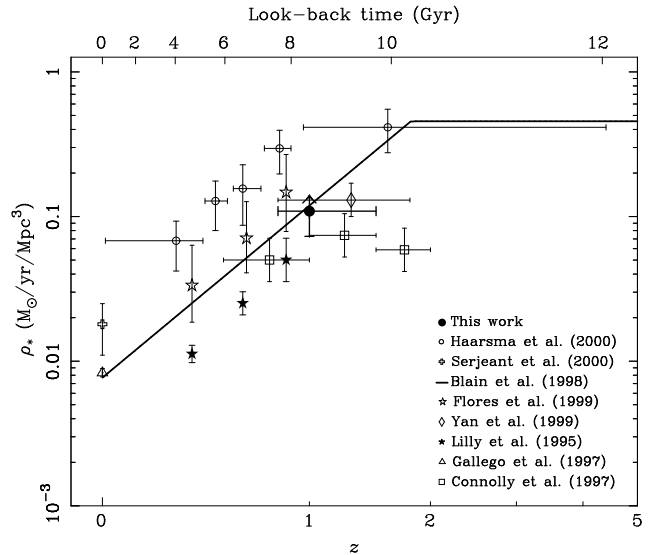
\caption{\small The evolution of the star formation density in the
Universe as measured using a range of long-wavelength tracers (which
should be relatively insensitive to dust obscuration).  We plot our
measurement of the star formation density in the dusty ERO population,
the lower limit of the error-bar on this point corresponds to our
conservative estimate of the star formation density in this population
(see text).  We compare this to other estimates from the radio
(Haarsma et al.\ 2000; Serjeant et al.\ 2000);
far-infrared/submillimeter (Blain et al.\ 1999a); mid-infrared (Flores
et al.\ 1999); near-infrared (Connolly et al.\ 1997), H$\alpha$
(Gallego et al.\ 1997; Yan et al.\ 1999) and UV/optical (Lilly et al.\
1995). }
\end{inlinefigure}

Hence we estimate cumulative restframe 1.4-GHz luminosities from star
formation in the two subsamples of $L_{1.4} > (8.0\pm 1.3)\times
10^{24}$\,W\,Hz and $L_{1.4} = (12.2\pm 1.5)\times 10^{24}$\,W\,Hz
respectively.  These assume a K-correction based on a $\nu^{-0.7}$
spectrum typical of star forming galaxies (Richards 1998), which gives
a correction of $1.62^{+0.28}_{-0.12}$, where the error is dominated
by the redshift range of the galaxies.  We now use the local
correlation between 1.4-GHz luminosity and far-infrared luminosity
(Condon et al.\ 1991) to convert this estimate into a cumulative
far-infrared luminosity for the sample, we note that the validity of
this correlation has recently been confirmed at the redshifts studied
here by Garrett (2002).  In this way we obtain cumulative far-infrared
luminosities of: $L_{FIR} >(16.1\pm 2.5)\times 10^{12} L_\odot$ and
$(24.3\pm 3.7)\times 10^{12} L_\odot $ respectively.  We estimate that
these correspond to SFRs for $M\geq 5$M$_\odot$ of $>1450\pm
220$\,M$_\odot$\,yr$^{-1}$ and $2180\pm 330$\,M$_\odot$\,yr$^{-1}$
using the conversion in Condon (1992).  If we instead adopt the new
calibration of Yun, Reddy \& Condon (2001), we would estimate star
formation rates roughly half of the quoted values (see also Carilli et
al.\ 2002).  However, for comparison with previous work we retain the
Condon (1992) calibration.  Assuming a Salpeter IMF from
0.1--100\,M$_\odot$ these rates are $\sim 5.5\times$ higher.

To calculate the volume probed by our ERO sample we use the photometric
redshift estimates for the 20 radio-detected and undetected EROs which
are classed as dusty starbursts and find a quartile range of
$z=0.8$--1.5 (in reasonable agreement with the spectroscopically
determined redshift range from Cimatti et al.\ 2002, $z=0.8$--1.4, see
Daddi et al.\ 2002), which gives a corresponding volume of $(1.1\pm
0.2) \times 10^5$\,Mpc$^3$ at a median redshift of $z=1$. The quoted
uncertainty corresponds to changing either of the bounds on the volume
by $\delta z = \pm 0.2$ and encompases the estimate we would derive if
we had instead adopted the redshift range for EROs from Daddi et al.\
(2002).

Combining this volume with the star formation rates we estimate
equivalent star formation densities of
$\stackrel{.}{\rho_*}($0.1--100\,M$_\odot)>0.073\pm
0.018$\,M$_\odot$\,yr$^{-1}$\,Mpc$^{-3}$ and $0.11\pm
0.03$\,M$_\odot$\,yr$^{-1}$\,Mpc$^{-3}$ (if the AGN contamination is
predominantly in the most luminous, unresolved radio sources the latter
estimate will fall to $\sim 0.09$\,M$_\odot$\,yr$^{-1}$\,Mpc$^{-3}$).
We compare these measurements to other estimates of the star formation
density in Fig.~7.

Our measurement of the star formation density in the radio-detected ERO
population at $z\sim 1$ is comparable to estimates of that in
H$\alpha$-selected galaxies at these epochs (Yan et al.\ 1999), and
greater than that seen in optically-selected galaxies (Lilly et al.\
1995; Connolly et al.\ 1997).  There is little or no overlap between
the ERO population used in our analysis and the classes of galaxies
which are employed in these previous studies and hence their
contributions to the total star formation density are distinct.  This
implies that the activity in obscured galaxies makes a sizable
contribution to the total star formation at $z\sim 1$, especially when
allowance is made for the proportion of the obscured star forming
population which are missing from our ERO sample due to the escape of a
small fraction of UV light.  The significant contribution from obscured
star formation at $z\sim 1$ confirms earlier claims by Flores et al.\
(1999) and Haarsma et al.\ (2001) and is particularly important as this
era represents the peak in the star formation activity in the Universe
(note that when integrated over cosmic time, the distribution in Fig.~7
is highly peaked at $z\sim 1$--2).  This result therefore extends the
conclusions of the mid-infrared, submillimeter and radio surveys, that
the use of obscuration-independent measures of the star formation rate
is critical to reliably estimate the total star formation density, from
$z\sim 2$--3 to the era when most of the stars seen in galaxies today
were formed.

\subsection{Dusty EROs in the Far-infrared and Submillimeter}

We now estimate the volume density of the most strongly star forming
EROs.  Based on the photometric redshifts for the 11 radio-detected
EROs which are well-fit by a dusty SED, we estimate a median radio
luminosity of $L_{1.4} =10^{23.6\pm 0.3}$\,W\,Hz$^{-1}$ (compared
to the Galaxy's radio luminosity of $L_{1.4}
=10^{21.4}$\,W\,Hz$^{-1}$).  Assuming these galaxies follow the local
radio--far-infrared correlation, this translates into a median
far-infrared luminosity of $L_{FIR}/L_\odot =10^{11.9\pm 0.3}$
for the subsample.  The brightest five EROs in this subsample have
inferred far-infrared luminosities above $L_{FIR}\geq 10^{12}L_\odot$
which classes them as ultraluminous infrared galaxies (ULIRGs).  Based
on these five galaxies we estimate a surface density of radio-detected
ULIRG EROs of $0.07\pm 0.03$\,arcmin$^{-2}$. This estimate could be
increased by $1.4\times$ to account for radio-detected EROs which are
not well-fit by our model SEDs, although equally any increases should
be offset against any AGN contribution to the radio luminosities of
these galaxies.

Using the volume containing the dusty ERO sample given in \S4.2 we
estimate an equivalent space density of ULIRG EROs as $\sim (0.5\pm
0.2)\times 10^{-4}$\,Mpc$^{-3}$ (and $\sim (2.7\pm 0.4)\times
10^{-4}$\,Mpc$^{-3}$ for all dusty EROs brighter than $K<20.5$).  This
is significantly higher than the space density of comparable
luminosity events locally, $0.03 \times 10^{-4}$\,Mpc$^{-3}$ (Sanders
\& Mirabel 1996), even before allowance is made for the fact that some
high-$z$ ULIRGs will be missing from our ERO selection due to their
blue colors (Trentham et al.\ 1999; Goldader et al.\ 2002).  Based
upon the far-UV colors of local ULIRGs it appears that more than half
of this population would be too blue at $z\sim 1$--2 to be included in
our ERO sample. Hence, if local ULIRGs are good analogs for the events
which produce the high-$z$ ULIRG population then our volume density of
$z\sim 1$ ULIRGs would require an upwards revision by a factor of $\gs
2\times$.  We conclude that there has been {\it at least} an order of
magnitude decrease in the number density of these luminous infrared
events between $z\sim 1$--2 (8--10\,Gyrs ago) and the present-day.
This trend is consistent with the evolutionary models of Blain et al.\
(1999a), which also fit the number counts of SCUBA galaxies at higher
redshifts (Fig.~7).

We can also estimate the submillimeter fluxes from the radio-detected
EROs and hence investigate their relationship to the 850-$\mu$m
selected samples detected by SCUBA.  Based on the individual radio
fluxes and estimated redshifts for the EROs we would predict a median
850-$\mu$m flux of 2\,mJy and a range of 0.3--5\,mJy (Carilli \& Yun
2000).  This places the bulk of these galaxies below the flux
limit of typical SCUBA observations of faint radio sources (e.g.\
Chapman et al.\ 2001) and at least half of them below the confusion
limit of the deepest (unlensed) SCUBA surveys.  Clearly the radio/ERO
selection allows us to probe the star forming population to lower star
formation rates (and also lower redshifts) than typically achieved in
submm surveys as demonstrated by our detection of all three cataloged
submm sources in our field (see Ledlow et al.\ 2002).

Nevertheless, the most luminous radio-detected EROs should overlap
with samples of ERO counterparts to SCUBA galaxies.  We can test
this by comparing the surface densities of these two populations.  Our
estimate of the surface density of $\gs 2$\,mJy radio-detected EROs is
$0.07\pm 0.03$\,arcmin$^{-2}$ at $K\leq 20.5$.  This can be compared
to the claim that $13\pm 9$\% of the counterparts to SCUBA galaxies
brighter than 2\,mJy have $K\leq 20.5$ and $(R-K)\geq 5.3$, equivalent
to a surface density of $\sim 0.11\pm 0.09$ arcmin$^{-2}$, (Smail et
al.\ 1999a, 2002; see also Mohan et al.\ 2002) and the estimate
surface density of EROs brighter than 8\,mJy of 0.03\,arcmin$^{-2}$
from Ivison et al.\ (2002).  Allowing for the large uncertainties in
these estimates, and the possible upward revision of the density of
radio-selected ULIRGs to account for badly-fit SEDs (including the
known SCUBA ERO \#01206), we conclude that they are in reasonable
agreement.  This suggests that by combining sensitive radio and
submillimeter surveys it will be possible to track the evolution of
the star formation density in highly-obscured galaxies from $z=0$ to
$z\gs 3$--5 (Owen et al.\ 2002; Chapman et al.\ 2002b).

Finally, we estimate the contribution from the dusty ERO population to the
far-infrared background at a wavelength corresponding to the peak of
their dust emission: $\lambda = (1+z) \lambda_{peak} \sim 120\mu$m.
Scaling from the 1.4-GHz emission we estimate that the dusty ERO
population produces $\nu I_\nu \sim 1 \times
10^{-9}$\,W\,m$^{-2}$\,sr$^{-1}$ at $\sim 100\mu$m, compared to
estimates of the background at this wavelength of $\nu I_\nu = (2\pm
1) \times 10^{-8}$\,W\,m$^{-2}$\,sr$^{-1}$ (Hauser et al.\ 1998;
Lagache \& Puget 2000).  Thus the dusty EROs could contribute $\sim 5$--10\%
of the far-infrared background at its peak at $\sim 100\mu$m.  This
compares to the SCUBA population which produces $\sim 100$\% of the
background at 850$\mu$m (Blain et al.\ 1999b; Smail et al.\ 2002;
Cowie et al.\ 2002) and probably the bulk of the far-infrared background
at $\lambda \gs 200\mu$m (where the dust emission peaks for a
population with a characteristic temperature of 40\,K at $z\sim
2$--3).  Both the dusty ERO and the SCUBA population contribute $\ll
1$\% of the extragalactic background at $1\mu$m, underlining the very
different nature of the populations selected in the optical and the
submillimeter/radio wavebands.

\subsection{The Passive ERO Population}

Our suggestion that at least a half of the faint ERO population are
dust-enshrouded starbursts obviously reduces the estimated volume
density of passive, evolved EROs.  We estimate that passive EROs at
$z\gs 1$ have a volume density of $\sim 3.6\pm0.6 \times
10^{-4}$\,Mpc$^{-3}$ for galaxies with a typical luminosity of
$M_V\sim -20.6\pm 0.4$. Passive evolution of the stellar populations would
produce $\sim 0.9$\,mags of brightening in the observed $K$-band to this
redshift, meaning that these galaxies would have $M_V\sim -19.7\pm
0.4$ at the present day if they evolve passively.  This assumes that
the passive EROs inhabit the same volume estimated for the dusty ERO
population, although as we discussed in \S4.1 these systems may lie in
a smaller volume and hence this volume density should be taken as a
lower limit.  In this regard we note that the $K<19.2$ passive sample
from Cimatti et al.\ (2002) have $z\sim 1.04\pm 0.15$, where the error
represents the 1-$\sigma$ dispersion, and a quartile range of
$z=0.8$--1.2, somewhat narrower than the redshift range we use -- and
suggesting a volume for the sample closer to $0.6\times
10^{5}$\,Mpc$^3$. Adopting this volume would roughly double our
estimate of the volume density of passive EROs.

The most frequently proposed local analogs to passive EROs are
elliptical galaxies, although it is clear that apparently passive EROs
embrace a wider cross-section of the $z\sim 1$ population than just
disk-less bulges (Smith et al.\ 2002b).  It is consequently difficult
to compare the volume density of this population with that of their
possible descendants at the present-day.  However, we note that the
combination of our reduction in the proportion of passive galaxies in
the ERO population, coupled with the increase in the diversity of
their morphological counterparts identified by Smith et al.\ (2002b),
means that previously well-fitting pure luminosity evolution models
with a high redshift of formation, $z>2$, will now over predict the
ERO counts at $K\sim 20$ (Daddi et al.\ 2000; Firth et al.\ 2001;
Smith et al.\ 2002a).  This suggests there may be some scope for the
recent formation (or transformation) of some fraction of the local
elliptical galaxy population (e.g.\ Aguerri \& Trujillo 2002).
Nevertheless, the reduction in the proportion of passive systems in the ERO
population is still not sufficient to reconcile the observed surface
density of evolved EROs with that predicted by semi-analytic galaxy
formation models (Smith et al.\ 2002a; Cole et al.\ 2000).

The proposed reduction in the passive ERO fraction also has ramifications for
several previous results based on the properties of the whole ERO
population.  The most obvious is the interpretation of the strong
clustering signal seen in ERO samples (Daddi et al.\ 2001, 2002;
McCarthy et al.\ 2001).  The strong angular clustering of the ERO
population, especially for the brightest examples, appears to
vindicate the predictions of hierarchical galaxy formation models
assuming these galaxies represent the passive, evolved progenitors of
present-day ellipticals (Firth et al.\ 2001).  In the same manner, the
apparent agreement between the angular clustering of
Lyman-break objects at $z\sim 3$ and theoretical models of halo
clustering has provided support for the suggestion that these objects are
the progenitors of massive ellipticals (Giavalisco \& Dickinson 2001).

However, the presence of a large proportion of, perhaps weakly
clustered, dusty EROs in the bright ERO population would tend to
dilute the angular clustering signal arising from the strongly
clustered passive EROs, unless there is a strong cross-correlation
between the two populations (see Daddi et al.\ 2002 and \S3.3).
Indeed, a further element comes into play when attempting to interpret
the projected clustering of ERO samples: the redshift depth of the
volume containing the galaxies.  If the passive ERO population lies in
a relatively narrow redshift range (see the toy models in \S4.1; Firth
et al.\ 2001) then the dilution effect from the depth of the sample
will be less than anticipated, requiring an intrinsically less
strongly clustered population to produced a given angular clustering
signal (Daddi et al.\ 2001).  Thus the effects of a narrower redshift
depth for the evolved ERO subsample and the increased proportion of
weakly-clustered, but typically faint, dusty EROs may cancel each other
out leaving the estimated correlation length for the bright, evolved
EROs close to that required for the progenitors of the $L^\ast$
elliptical galaxies seen in the local Universe, as estimated by Daddi
et al.\ (2001) and McCarthy et al.\ (2001).  We conclude that the
complex interplay between the sample volume and the mix of evolved and
obscured galaxies in the ERO population as a function of apparent
magnitude means that the interpretation of the angular clustering of
this class of galaxies is far from trivial.

\section{Conclusions}

The main conclusions of this work are:

\noindent{1)} We catalog 68 EROs within a 72.8 arcmin$^2$ field which
has the deepest 1.4-GHz observations of any region on the sky.  We find
that 31\% of these EROs are detected in the radio map down to a flux
limit of 12.6\,$\mu$Jy, of these radio-detected EROs, 33\% are resolved
in our VLA map with a typical FWHM of $1.4\pm 0.4''$ -- equivalent to
$12\pm 3$\,kpc at $z\sim 1$.  Moreover, the radio emission in these
EROs mirrors their $R$-band morphologies, suggesting that star
formation is responsible for the bulk of the radio flux.

\noindent{2)} Using our deep $RIzJHK$ imaging we photometrically
classify the radio-detected and undetected EROs into dusty or evolved
systems based on simple SED models.  We first analyse the colors of
those radio-detected with high quality photometry and conclude that the
majority of these are adequately described by a highly dust-reddened
starburst spectrum.  This provides support for applying this analysis
to the radio-undetected EROs to search for similar obscured, active
galaxies.

\noindent{3)} Using our simple photometric models we estimate that more
than 32\% of the radio-undetected EROs with high quality photometry, or
a minimum of 20\% of all radio-undetected EROs, have the colors of
dusty starbursts, similar to the radio-detected EROs.  We estimate the
rate of mis-classification of EROs in our analysis assuming that most of
the radio-detected EROs are actually dusty starbursts, and based on
this propose that $\sim 45$\% of the EROs with radio fluxes
$<12.6\mu$Jy have dusty SEDs.  ~From the continuity of the number
counts of radio-detected EROs we would expect that most of these
sources would have 1.4-GHz fluxes around 5\,$\mu$Jy.  We also find that
one of the 68 EROs has an SED consistent with galactic star, while
another apparent ERO appears to be a transient source.

\noindent{4)} Combining the samples of radio-detected and undetected
EROs which are photometrically classified as dusty we place a firm
lower limit of $\geq 30$\% on the proportion of dusty starbursts in the
whole ERO population.  To obtain a more realistic estimate we adopt a
correction for mis-classification and assume the resulting proportions
of dusty and evolved EROs are representative of the full sample (not
just those with the best photometry) we estimate that at least
45--50\%, and possibly up to $\sim 60\pm15$\%, of the {\it whole} ERO
population with $(R-K)\geq 5.3$ and $K\leq 20.5$ are probably dusty,
active galaxies.  Our fitting procedure also suggests a median redshift
of $z\sim 1.0\pm 0.3$ with a quartile range of $z=0.8$--1.5.  This is
in reasonable agreement with the spectroscopic survey of somewhat
brighter and bluer EROs by Cimatti et al.\ (2002).  Assuming that AGN
do not dominate the radio emission from these galaxies, we estimate a
volume density of EROs with far-infrared luminosities of $\gs 10^{12}
L_\odot$ (i.e.\ ULIRGs) of $0.5\times 10^{-4}$\,Mpc$^{-3}$, this is
over an order of magnitude higher than for equivalent luminosity events
at $z=0$ and confirms the strong evolution of the most dusty, active
galaxies out to $z\gs 1$ (Blain et al.\ 1999a).

\noindent{5)} We use the local far-infrared--radio correlation to
estimate star formation rates from the radio fluxes of the dusty EROs,
assuming that their radio emission is purely powered by massive star
formation.  Combining these estimates with the probable volume spanned
by our dusty ERO subsample we can calculate the star formation density
in this obscured population.  Adopting conservative assumptions we find
that the star formation density in star forming EROs at $z\sim 1$
probably exceeds $\stackrel{.}{\rho_*}($0.1--100\,M$_\odot)>0.073\pm
0.018$\,M$_\odot$\,yr$^{-1}$\,Mpc$^{-3}$ and maybe as high as
$\stackrel{.}{\rho_*}=0.11\pm 0.03$\,M$_\odot$\,yr$^{-1}$\,Mpc$^{-3}$.
This measurement is comparable to estimates of the star formation
density in H$\alpha$-selected galaxies at these epochs, and greater
than that seen in optically--selected galaxies, showing that the
activity in obscured systems may make a significant contribution to the
total star formation at $z\sim 1$.  We also estimate that the dusty ERO
population may produce up to 5--10\% of the far-infrared background at
its peak at $\sim 100\mu$m, underlining the cosmological significance
of these events for understanding the history of star formation in the
Universe.

\noindent{6)} We propose that the apparent break in the integrated
counts of the EROs at $K\sim 19$--20 reflects the peak of the
contribution from passive, evolved galaxies to this population.  At
magnitudes fainter than $K\sim 20$ we predict that the population will
be increasing dominated by dusty, active galaxies.  The continuity in
the properties of these faint dusty EROs connects them to samples of
submm-selected obscured starbursts at higher redshifts and fainter
$K$-band magnitudes.  This proposal can be tested by studies of the
multiwavelength properties of extremely faint samples of EROs, $K\sim
23$, and their relationship to the blank-field populations of radio
sources.

\noindent{7)} By identifying the likely contribution from dusty EROs to
the total ERO population we estimate that the volume density of
passive, evolved galaxies brighter than $K\leq 20.5$ is $\gs 4\times
10^{-4}$\,Mpc$^{-3}$. Recent results point to a morphological diversity
in the passive ERO population which would suggest that these galaxies
should not be simply viewed as passively evolving ellipticals (Smith et
al.\ 2002b).  We conclude that previously well-fitting pure luminosity
evolution models, which described the number counts of the whole ERO
sample as due to a passively evolving elliptical population formed at
high redshifts, will probably over predict the number of truly passive
EROs and hence there is scope for the recent (trans)formation of some
local elliptical galaxies.  Nevertheless, the presence of luminous,
passive galaxies at $z\sim 1$ does point to a phase of massive spheroid
formation at higher redshifts, consistent with the interpretation of
the SCUBA population as the formation phase of massive,
proto-ellipticals (Lilly et al.\ 1999; Smail et al.\ 2002, 2003).  The
typical, reddening-corrected luminosities of the radio-detected ERO
population at $z\sim 1$ suggests that these galaxies will evolve into
sub-$L^\ast$ systems by the present-day unless their star formation
activity continues at a high rate for $\gg 10^8$\,yrs.

The area analysed in this work covers a small fraction of
the sensitive field of our VLA map -- we expect to be able to
quadruple the ERO sample available for analysis in the near
future.  In addition we are undertaking a full photometric analysis of
the $\sim 1500$ optically-bright $\mu$Jy radio sources in this field
using our panoramic $U\!BV\!RIzJHK$ imaging dataset (Owen et al.\
2002), constrained by deep spectroscopy from GMOS on Gemini.

\section*{Acknowledgements}

We thank Taddy Kodama, F.\ Nakata and S.\ Okamura for kindly allowing
us to use their exquisite Subaru Suprime-Cam imaging.  We also
acknowledge useful conversations with Dave Alexander, 
Andrew Blain, Scott Chapman,
Andrea Cimatti, Adriano Fontana, Bianca Poggianti, Alice Shapley,
Graham Smith and Chuck Steidel.  We thank an anonymous referee for
their thorough and constructive comments which improved the clarity and
presentation of this work.  IRS acknowledges support from a Royal
Society University Research Fellowship and a Philip Leverhulme Prize
Fellowship.  GEM gratefully acknowledges the support of JPL (contracts
\#1147 and \#1166) and also the support of Vanguard Research, Inc.

%
%

\setcounter{table}{0}
\begin{table*}
{\scriptsize
\begin{center}
\caption{\hfil ERO Catalogue \hfil}
\hspace*{-1.cm}
\begin{tabular}{rcccrcccccccl}
\hline\hline
ID    &   R.A./Dec.\  & $K_{tot}$ & $(R-K)$ & $S_{1.4}$~ & $R_{ap}^a$ & $I_{ap}$ & $z_{ap}$ & $J_{ap}$ & $H_{ap}$ & $K_{ap}$ & SED$^b$ & Morph$^c$ \cr
      &     (J2000)   &           &         & ($\mu$Jy) &  \cr
\hline
\noalign{\smallskip}
01045 &  09 42 29.260~ 47 00 31.30 & 19.10 & 5.31  & $ 37.5$ & $24.68[03]$ & $23.50[06]$ & $22.86[11]$ & $21.27[13]$ & $20.16[08]$ & $19.37[07]$ & B & .../C\cr
01049 &  09 42 29.750~ 47 03 22.50 & 18.72 & 5.53  & $ 44.4$ & $25.00[05]$ & $24.01[07]$ & $23.12[12]$ & $21.57[16]$ & $20.57[11]$ & $19.47[07]$ & D & .../I\cr
01050 &  09 42 30.500~ 46 56 41.70 & 19.88 & 5.74  & $ 39.4$ & $25.78[08]$ & $24.37[09]$ & $23.28[13]$ & $22.07[26]$ & $20.91[17]$ & $20.04[11]$ & D & .../I\cr
01053 &  09 42 30.610~ 47 00 21.70 & 17.51 & 5.60  & $195.6$ & $23.65[01]$ & $22.28[01]$ & $21.18[03]$ & $20.32[07]$ & $19.17[04]$ & $18.05[03]$ & D & .../I\cr
01071 &  09 42 32.940~ 46 57 31.70 & 18.97 & $>6.01$  & $112.2$ & $>26.7$ & $25.40[30]$ & $>24.1$ & $>22.7$ & $21.71[27]$ & $20.69[17]$ & D & .../F\cr
01081 &  09 42 34.140~ 47 00 14.40 & 20.20 & 5.44  & $ 70.2$ & $25.81[13]$ & $24.83[12]$ & $>24.1$ & $22.20[24]$ & $21.84[31]$ & $20.37[13]$ & D & .../I\cr
01105 &  09 42 37.980~ 47 01 20.90 & 19.33 & 5.88  & $ 51.9$ & $25.78[09]$ & $24.06[07]$ & $23.45[17]$ & $21.84[20]$ & $20.36[10]$ & $19.91[10]$ & B & .../I\cr
01110 &  09 42 38.880~ 47 01 33.60 & 19.95 & 6.13  & $ 45.7$ & $26.07[14]$ & $24.59[11]$ & $23.89[33]$ & $>22.7$ & $21.26[23]$ & $19.94[10]$ & D & .../I\cr
01115 &  09 42 40.000~ 47 02 28.10 & 19.89 & 5.45  & $ 44.1$ & $25.84[11]$ & $24.57[12]$ & $23.88[26]$ & $22.28[23]$ & $22.02[31]$ & $20.39[12]$ & D & .../I\cr
01118 &  09 42 40.340~ 47 00 16.30 & 18.88 & 5.32  & $ 16.0$ & $24.71[03]$ & $23.55[04]$ & $22.51[07]$ & $21.33[16]$ & $20.53[13]$ & $19.39[08]$ & D & I/I\cr
01137 &  09 42 43.430~ 46 56 23.00 & 19.62 & 5.30  & $ 24.4$ & $25.63[10]$ & $24.34[09]$ & $24.10[31]$ & $22.13[21]$ & $20.91[12]$ & $20.33[12]$ & B & I/I\cr
01150 &  09 42 44.840~ 46 56 37.50 & 19.66 & $>6.05$ & $ 34.4$ & $>26.7$ & $>25.6$ & $>24.1$ & $>22.7$ & $21.86[30]$ & $20.65[16]$ & ... & F/F\cr
01157 &  09 42 46.250~ 46 59 30.20 & 20.07 & $>6.55$ & $ 64.5$ & $>26.7$ & $>25.6$ & $24.16[33]$ & $>22.7$ & $>21.9$ & $20.15[10]$ & ... & I/F \cr
01176 &  09 42 49.970~ 46 57 40.60 & 20.08 & 5.66  & $ 32.9$ & $26.08[12]$ & $24.52[11]$ & $>24.1$ & $>22.7$ & $20.91[17]$ & $20.41[14]$ & D & I/I\cr
01206 &  09 42 54.540~ 46 58 44.10 & 20.10 & $>5.65$  & $ 81.1$ & $>26.0$ & $>25.6$ & $>24.1$ & $>22.7$ & $>21.9$ & $20.35[20]$ & ... & I$^d$/F \cr
01273 &  09 43 02.680~ 47 01 42.60 & 20.05 & 5.97  & $199.9$ & $26.64[18]$ & $25.48[23]$ & $>24.1$ & $>22.7$ & $>21.9$ & $20.66[17]$ & D & F/F \cr
01338 &  09 43 09.750~ 47 00 04.10 & 20.03 & $>6.02$  & $ 23.2$ & $>26.7$ & $>25.6$ & $>24.1$ & $>22.7$ & $>21.9$ & $20.68[21]$ & ... & I/F\cr
11018 &  09 43 13.067~ 46 55 50.07 & 18.59 & 5.94  & $  17.0$ & $25.34[16]$ & $23.76[08]$ & $22.96[13]$ & $22.62[31]$ & $21.14[17]$ & $19.40[06]$ & B & I/I\cr
11077 &  09 43 00.187~ 46 56 01.51 & 19.10 & 5.87  & $  24.6$ & $25.93[14]$ & $25.49[24]$ & $>24.1$ & $>22.7$ & $21.34[21]$ & $20.06[12]$ & E & I/I\cr
12861 &  09 42 58.729~ 47 03 39.62 & 19.60 & 6.73  & $  14.1$ & $26.54[18]$ & $24.56[14]$ & $23.98[32]$ & $22.06[24]$ & $20.70[14]$ & $19.81[09]$ & B & .../I\cr
13963 &  09 43 13.663~ 47 02 07.60 & 20.45 & $>6.23$  & $  16.3$ & $>26.7$ & $25.02[17]$ & $>24.1$ & $>22.7$ & $21.65[30]$ & $20.47[15]$ & D & I/I\cr
\noalign{\smallskip}
 1091 &  09 42 27.721~ 46 56 07.02 & 19.47 & $>6.30$  & $  <12.6$ & $>26.7$ & $>25.6$ & $>24.1$ & $>22.7$ & $21.75[33]$ & $20.40[16]$ & ... & .../F\cr
 1097 &  09 42 30.347~ 46 56 05.44 & 20.15 & 5.34  & $  <12.6$ & $25.59[07]$ & $24.46[08]$ & $23.43[17]$ & $22.58[33]$ & $>21.9$ & $20.25[13]$ & B & .../I\cr
 1136 &  09 42 40.436~ 46 56 13.55 & 19.69 & 5.56  & $  <12.6$ & $25.74[17]$ & $24.62[19]$ & $23.54[23]$ & $>22.7$ & $21.36[28]$ & $20.18[12]$ & D & F/C\cr
 1183 &  09 42 55.042~ 46 56 26.84 & 19.98 & 5.31  & $  <12.6$ & $25.54[07]$ & $25.82[28]$ & $>24.1$ & $>22.6$ & $21.12[18]$ & $20.23[11]$ & E & I/I\cr
 1188 &  09 42 56.197~ 46 56 24.13 & 19.46 & $>6.28$  & $  <12.6$ & $>26.7$ & $>25.6$ & $>24.1$ & $>22.7$ & $>21.9$ & $20.42[14]$ & ... & F/F\cr
 1200 &  09 42 52.459~ 46 56 31.11 & 18.71 & $>6.94$  & $  <12.6$ & $>26.1$ & $24.27[33]$ & $>23.4$ & $21.08[12]$ & $19.90[08]$ & $19.16[06]$ & B & C/I\cr
 1206 &  09 43 04.291~ 46 56 32.52 & 19.91 & $>6.09$  & $  <12.6$ & $>26.7$ & $>25.6$ & $>24.1$ & $>22.7$ & $>21.9$ & $20.61[17]$ & ... & F/F\cr
 1321 &  09 43 01.416~ 46 56 59.37 & 19.52 & 6.15  & $  <12.6$ & $26.27[19]$ & $25.44[29]$ & $>24.1$ & $>22.7$ & $20.91[16]$ & $20.13[10]$ & E & I/I\cr
 1349 &  09 42 36.967~ 46 57 06.70 & 19.55 & 5.54  & $  <12.6$ & $26.04[15]$ & $24.32[11]$ & $23.31[14]$ & $21.74[17]$ & $21.68[30]$ & $20.50[15]$ & E & .../I\cr
 1473 &  09 42 54.991~ 46 57 37.60 & 19.35 & 5.66  & $  <12.6$ & $25.20[10]$ & $23.26[05]$ & $22.50[08]$ & $21.12[12]$ & $20.35[11]$ & $19.54[07]$ & B & C/C\cr
 1523 &  09 42 55.919~ 46 57 46.88 & 19.57 & 5.32  & $  <12.6$ & $24.79[09]$ & $23.08[07]$ & $22.32[08]$ & $21.48[17]$ & $20.54[14]$ & $19.47[08]$ & D & C$^f$/C\cr
 1571 &  09 42 49.721~ 46 57 54.12 & 19.92 & $>5.82$  & $  <12.6$ & $>26.7$ & $>25.6$ & $>24.1$ & $>22.7$ & $>21.9$ & $20.88[21]$ & ... & F/F\cr
 1574 &  09 42 53.270~ 46 57 56.70 & 19.50 & $>6.59$  & $ <12.6$ & $>26.7$ & $24.29[10]$ & $23.54[18]$ & $22.05[24]$ & $21.29[22]$ & $20.11[12]$ & D & I/I\cr
 1578 &  09 42 48.720~ 46 58 00.70 & 19.91 & $>6.21$  & $  <12.6$ & $>26.7$ & $>25.6$ & $24.02[34]$ & $>22.7$ & $>21.9$ & $20.49[14]$ & ... & F/F\cr
 1762 &  09 42 35.264~ 46 58 38.27 & 19.56 & 5.30  & $  <12.6$ & $25.77[07]$ & $>25.6$ & $>24.1$ & $22.44[31]$ & $>21.9$ & $20.47[15]$ & E & .../F\cr
 1831 &  09 42 29.994~ 46 58 53.86 & 20.25 & $>5.87$  & $  <12.6$ & $>26.7$ & $>25.6$ & $>24.1$ & $>22.7$ & $21.15[18]$ & $20.83[21]$ & ... & .../F\cr
 1931 &  09 42 59.183~ 46 59 17.34 & 18.68 & $>6.91$  & $  <12.6$ & $>25.9$ & $24.03[16]$ & $23.46[27]$ & $21.05[12]$ & $19.80[08]$ & $18.99[05]$ & B & I$^e$/I\cr
 1961 &  09 42 27.404~ 46 59 19.52 & 20.33 & $>5.77$  & $  <12.6$ & $>26.7$ & $>25.6$ & $>24.1$ & $>22.7$ & $>21.9$ & $20.93[23]$ & ... & .../F\cr
 1998 &  09 42 46.730~ 46 59 29.12 & 19.78 & 5.53  & $  <12.6$ & $25.41[06]$ & $23.83[06]$ & $22.95[09]$ & $>22.7$ & $21.30[19]$ & $19.88[08]$ & D & C/I\cr
 2047 &  09 42 45.795~ 46 59 36.12 & 20.26 & 5.85  & $  <12.6$ & $26.39[13]$ & $24.82[15]$ & $>24.1$ & $22.17[22]$ & $>21.9$ & $20.54[16]$ & E & F/I\cr
 2128 &  09 43 12.079~ 46 59 58.69 & 18.86 & 5.39  & $  <12.6$ & $25.63[16]$ & $24.73[19]$ & $>24.1$ & $22.83[34]$ & $21.16[17]$ & $20.25[12]$ & D & I/F\cr
 2176 &  09 43 12.745~ 47 00 02.55 & 19.85 & 6.03  & $  <12.6$ & $26.24[26]$ & $25.10[26]$ & $24.06[32]$ & $22.75[34]$ & $21.77[31]$ & $20.21[15]$ & D & C/I\cr
 2334 &  09 42 36.727~ 47 00 27.71 & 20.21 & $>5.89$  & $  <12.6$ & $>26.7$ & $>25.6$ & $>24.1$ & $>22.7$ & $>21.9$ & $20.81[21]$ & ... & .../F\cr
 2351 &  09 42 23.653~ 47 00 33.41 & 19.85 & $>6.18$  & $  <12.6$ & $>26.7$ & $>25.6$ & $23.57[34]$ & $>22.7$ & $>21.9$ & $20.52[17]$ & ... & .../...\cr
 2405 &  09 42 34.421~ 47 00 44.61 & 20.29 & 5.51  & $  <12.6$ & $26.20[15]$ & $24.28[12]$ & $>24.1$ & $22.36[27]$ & $21.41[20]$ & $20.68[16]$ & E & .../I\cr
 2418$^g$ &  09 42 42.925~ 47 00 53.59 & 17.79 & 6.08  & $  <12.6$ & $24.20[03]$ & $20.95[01]$ & $19.82[01]$ & $18.95[03]$ & $18.37[03]$ & $18.12[03]$ & B & .../C\cr
 2510 &  09 42 38.542~ 47 01 11.67 & 19.14 & 5.37  & $  <12.6$ & $24.99[04]$ & $23.53[04]$ & $23.06[11]$ & $21.51[15]$ & $20.34[10]$ & $19.61[08]$ & B & .../I\cr
 2528 &  09 42 29.343~ 47 01 10.44 & 20.34 & $>5.55$  & $  <12.6$ & $>26.7$ & $>25.6$ & $>24.1$ & $>22.7$ & $>21.9$ & $21.15[28]$ & ... & .../F\cr
 2549$^h$ &  09 42 57.330~ 47 01 17.81 & 19.01 & $>7.61$  & $  <12.6$ & $>26.7$ & $>25.6$ & $>24.1$ & $19.31[04]$ & $18.00[03]$ & $19.09[06]$ & B & .../F\cr
 2593 &  09 42 30.669~ 47 01 27.41 & 19.55 & 5.92  & $  <12.6$ & $25.85[12]$ & $24.48[13]$ & $23.20[14]$ & $>22.7$ & $21.18[18]$ & $19.94[10]$ & D & .../I\cr
 2637 &  09 42 36.916~ 47 01 37.91 & 19.42 & 5.96  & $  <12.6$ & $25.79[10]$ & $24.15[10]$ & $24.02[34]$ & $21.54[16]$ & $20.67[13]$ & $19.83[09]$ & B & .../I\cr
 2762 &  09 43 07.026~ 47 03 08.20 & 20.11 & $>6.02$  & $  <12.6$ & $>26.7$ & $>25.6$ & $>24.1$ & $>22.7$ & $>21.9$ & $20.68[20]$ & ... & .../F\cr
 2769 &  09 43 01.067~ 47 02 59.08 & 19.57 & $>5.79$  & $  <12.6$ & $>26.7$ & $25.20[23]$ & $>24.1$ & $>22.7$ & $>21.9$ & $20.91[21]$ & ... & .../F\cr
 2770 &  09 43 00.967~ 47 03 16.07 & 19.65 & 5.90  & $  <12.6$ & $26.48[19]$ & $25.29[26]$ & $23.66[20]$ & $>22.7$ & $>21.9$ & $20.58[19]$ & D & .../I\cr
 2778 &  09 42 43.902~ 47 03 15.18 & 20.15 & $>5.98$  & $  <12.6$ & $>26.7$ & $>25.6$ & $>24.1$ & $>22.7$ & $>21.9$ & $20.72[20]$ & ... & .../F\cr
 2801 &  09 43 01.485~ 47 03 16.40 & 20.13 & $>5.35$  & $  <12.6$ & $>26.7$ & $>25.6$ & $>24.1$ & $>22.7$ & $>21.9$ & $21.35[32]$ & ... & .../F\cr
 2807 &  09 43 06.467~ 47 03 11.84 & 19.78 & $>5.49$  & $  <12.6$ & $>26.7$ & $>25.6$ & $24.15[32]$ & $>22.7$ & $>21.9$ & $21.21[28]$ & ... & .../F\cr
 2862 &  09 42 30.473~ 47 03 17.54 & 19.03 & 5.49  & $  <12.6$ & $25.36[06]$ & $24.41[09]$ & $22.88[10]$ & $21.99[23]$ & $20.59[11]$ & $19.87[09]$ & E & .../I\cr
 2972 &  09 42 26.348~ 47 03 45.41 & 19.39 & 5.79  & $  <12.6$ & $26.72[28]$ & $>25.6$ & $>24.1$ & $>22.7$ & $>21.9$ & $20.93[23]$ & ... & .../F\cr
 2976 &  09 42 28.057~ 47 04 00.26 & 19.14 & $>6.00$  & $  <12.6$ & $>26.7$ & $25.48[26]$ & $>24.1$ & $>22.7$ & $>21.9$ & $20.70[19]$ & ... & .../F\cr
 2980 &  09 42 37.738~ 47 03 49.80 & 19.70 & 5.37  & $  <12.6$ & $26.03[16]$ & $>25.6$ & $>24.1$ & $22.69[34]$ & $>21.9$ & $20.66[20]$ & E & .../I\cr
 2984 &  09 42 30.238~ 47 03 44.30 & 19.61 & 5.59  & $  <12.6$ & $25.98[12]$ & $24.71[13]$ & $>24.1$ & $21.84[20]$ & $21.44[24]$ & $20.39[14]$ & E & .../I\cr
 3004 &  09 42 55.478~ 47 04 14.86 & 19.44 & 6.17  & $  <12.6$ & $26.62[22]$ & $25.33[23]$ & $23.84[21]$ & $22.24[28]$ & $21.36[25]$ & $20.44[14]$ & E & .../I\cr
 3082 &  09 43 03.066~ 47 04 04.24 & 19.78 & 5.68  & $  <12.6$ & $26.30[14]$ & $25.92[33]$ & $24.10[28]$ & $22.09[24]$ & $>21.9$ & $20.61[15]$ & E & .../I\cr
 3732 &  09 42 37.383~ 47 03 32.32 & 18.59 & 5.62  & $  <12.6$ & $24.62[04]$ & $23.16[04]$ & $22.16[05]$ & $20.63[08]$ & $19.87[07]$ & $19.01[05]$ & B & .../C\cr
 3930 &  09 42 53.314~ 47 02 05.63 & 20.11 & 5.40  & $  <12.6$ & $25.90[09]$ & $25.06[19]$ & $>24.1$ & $>22.7$ & $20.97[15]$ & $20.50[12]$ & D & F/C\cr
 3942 &  09 42 27.216~ 47 02 09.18 & 19.31 & 5.79  & $  <12.6$ & $25.36[06]$ & $23.94[07]$ & $23.32[16]$ & $21.24[11]$ & $20.51[10]$ & $19.57[07]$ & B & .../I\cr
\hline
\noalign{\smallskip}
\end{tabular}
\end{center}

$a$) Photometric errors in 0.01 mags are listed in [].
$b$) Best-fitting model SED: D, dusty; E, evolved; B, badly fit.
$c$) Morphological classification in $K$- and $R$-bands: ``C'', compact/regular;
``I'', disky, disturbed or amorphous; ``F'', too faint to classify.
$d$) SMM\,J09429+4658 (Smail et al.\ 1999).
$e$) ERO \#333 (Smail et al.\ 1999).
$f$) ERO Cl\,0939+4713 A (Persson et al.\ 1993).
$g$) M star.
$h$) Transient source.
}
\end{table*}


\begin{thebibliography}{99}
\bibitem{} Aguerri J.A.L., Trujillo I., 2002, MNRAS, in press{}{}
\bibitem{} Alexander D.M., Brandt W.N., Hornschemeier A.E.,
           Garmire G.P., Schneider D.P., Bauer F.E., Griffiths R.E., 2001, AJ, 122, 2156 {}{}
\bibitem{} Alexander D.M., Vignali C., Bauer F.E., Brandt W.N., Hornschemeier A.E.,
           Garmire G.P., Schneider D.P., 2002, AJ, in press {}{}
\bibitem{} Barger A.J., Cowie L.L., Richards E., 2000, AJ, 119, 2092 {}{}
\bibitem{} Benn C.R., Rowan-Robinson M., McMahon R., Broadhurst T.J., Lawrence A.,
           1993, MNRAS, 263, 98{}{}
\bibitem{} Bertin E., Arnouts S., 1996, A\&A, 117, 393{}{}
\bibitem{} Blain A.W., Smail I., Ivison R.J., Kneib J.-P.,
           1999a, MNRAS, 302, 632{}{}
\bibitem{} Blain A.W., Kneib J.-P., Ivison R.J., Smail I., 1999b,
           ApJ, 512, L87{}{}
\bibitem{} Bolzonella M., Miralles J.-M., Pell\'o R., 2000, A\&A, 363, 476{}{}
\bibitem{} Brusa M., Comastri A., Daddi E., Cimatti A., Vignal C., in
           ``X-ray spectroscopy of AGN with {\it Chandra} and {\it XMM/Newton}'', 
           eds.\ Boller Th., Komossa S., Kahn S., Kunioda H., 2002.{}{}
\bibitem{} Calzetti, D., Armus, L., Bohlin, R.C., Kinney, A.L., et al., 2000, ApJ, 533, 682{}{}
\bibitem{} Carilli C.L., Yun M.S., 2000, ApJ, 530, 618 {}{}
\bibitem{} Carilli C.L., Poggianti B.M., et al., 2002, in prep{}{}
\bibitem{} Chapman, S.C., McCarthy P.J., Persson, S.E., 2000, AJ, 120, 1612{}{}
\bibitem{} Chapman S.C., Richards E.A., Lewis G., Wilson G., Barger A.,
           2001, ApJ, 548, L147{}{}
\bibitem{} Chapman S.C., Barger A.J., Cowie L.L., Borys C., et al.,
           2002a, ApJ, submitted{}{}
\bibitem{} Chapman S.C., Blain A.W., Ivison R.J., Smail I., 2002b,
           Nature, submitted{}{}
\bibitem{} Chen, H.-W., McCarthy, P.J., Marzske, R.O., Wilson, J., et al., 2002, ApJ, 570, 54{}{}
\bibitem{} Cimatti A., Daddi E., Mignoli M.,
           Pozzetti, L., Renzini A., et al., 2002, A\&A, 381, L68{}{}
\bibitem{} Cole S., Lacey C.G., Baugh C.M., Frenk C.S., 2000,
           MNRAS, 319, 204{}{}
\bibitem{} Cole S., Norberg P., Baugh C.M., Frenk C.S., et al., 2001, MNRAS, 326, 255{}{}
\bibitem{} Condon J.J., 1992, ARAA, 30, 575{}{}
\bibitem{} Condon J.J, Anderson M.L., Helou G., 1991, ApJ, 376, 95{}{}
\bibitem{} Connolly A., Szalay A.S., Dickinson M., Subbarao M.U., Brunner R.J.,
           1997, ApJ, 486, L11{}{}
\bibitem{} Cowie L.L., Barger A.J., Bautz M.W., Capak P., Crawford C.S.,
           Fabian A.C., Hu E.M., Iwamuro F., Kneib J.-P., Maihara T., Motohara K.,
           2001, ApJL, 551, 9{}{}
\bibitem{} Cowie L.L., Barger A.J., Kneib J.-P., 2002, AJ, in press{}{}
\bibitem{} Cram L., Hopkins A., Mobasher B., Rowan-Robinson M.,  1998, ApJ, 507, 155{}{}
\bibitem{} Daddi E., Cimatti A., Pozzetti L., Hoekstra H.,
           Rottgering H.J.A., Renzini A., Zamorani G., Mannucci F.,
           2000, A\&A, 361, 535{}{}
\bibitem{} Daddi E., Broadhurst T.J., Zamorani G., Cimatti A., Rottgering H.,
           Renzini A., 2001, A\&A, 376, 745{}{}
\bibitem{} Daddi E., Cimatti A.,  Broadhurst T.J., Renzini A., Zamorani G., et al.,
           2002, A\&A, in press{}{}
\bibitem{} Dey A., Graham J.R., Ivison R.J., Smail I.,
           Wright G.S., Liu M.C., 1999, ApJ, 519, 610{}{}
\bibitem{} Dressler A., Oemler A., Sparks W.B., Lucas R.A., 1994, ApJ, 430, 107{}{}
\bibitem{} Dressler A., Oemler A., Couch W.J., Smail I., Ellis
           R.S., Barger A., Butcher H., Poggianti B.M., Sharples R.M., 1997, ApJ, 490, 577
\bibitem{} Dressler A., et al., 2002, ApJ, in prep{}{}
\bibitem{} Flores H., Hammer F., Thuan T.X., et al., 1999, ApJ, 517, 148{}{}
\bibitem{} Firth A.E., Somerville R.S., McMahon R.G., Lahav O., et al., 2002, MNRAS, submitted{}{}
\bibitem{} Franceschini A., Fadda D., Cesarsky C., Elbaz D., Flores H.,
           Granato G.L., 2002, ApJ, submitted {}{}
\bibitem{} Gallego J., Zamorano J., Rego M., Vitores A.G., 1997, ApJ, 475, 502{}{}
\bibitem{} Gardner J.P., Cowie L.L., Wainscoat R.J., 1993, ApJL, 415, 9{}{}
\bibitem{} Gallimore J.F., Keel W.C., 1993, AJ, 106, 1337{}{}
\bibitem{} Gal-Yam A., Ofek E.O., Filippenko A.V., Chornock R., Li W., 2002, PASP, submitted{}{}
\bibitem{} Garrett M.A., 2002, A\&A, in press{}{}
\bibitem{} Georgakakis A.E., Mobasher B., Cram L., Hopkins A., 1999, MNRAS, 310, L15{}{}
\bibitem{} Giavalisco M., Dickinson M., 2001, ApJ, 550, 177{}{}
\bibitem{} Goldader J.D., Meurer G., Heckman T.M., Seibert M., Sanders D.B., et al., 2002, ApJ,
           in press{}{}
\bibitem{} Haarsma D.B., Partridge R.B., Windhorst R.A., Richards E.A.,
           2000, ApJ, 544, 641{}{}
\bibitem{} Hammer F.,  Crampton D., Lilly S.J., Le Fevre O., Kenet T., 1995, MNRAS, 276, 1085{}{}
\bibitem{} Hauser M.G., Arendt R.G., Kelsall T., Dwek E., Edegard N., et al., 1998, ApJ, 508, 25{}{}
\bibitem{} Hawarden T.G., Leggett S.K., Letawsky M.B., Ballantyne D.R.,
           Casali M.M., 2001, MNRAS, 325, 563{}{}
\bibitem{} Hopkins A., Connolly A.J., Haarsma D.B., Cram L.E., 2001, AJ, 122, 288{}{}
\bibitem{} Iye M., Iwamuro F., Maihara T., Miyazaki S., Okamura S., et al., 2000, PASJ, 52, 9{}{}
\bibitem{} Ivison, R.J., Greve, T., Smail, I., Dunlop, J.S., et al., 2002, MNRAS, submitted{}{}
\bibitem{} Kodama T., Smail I., Nakata F., Okamura S.,
           Bower R.G., 2001, ApJL, 562, L9{}{}
\bibitem{} Lagache G., Puget J.L., 2000, A\&A, 355, L17{}{}
\bibitem{} Ledlow M., Smail I., Owen F.N., Keel W.C., Ivison R.J.,
           Morrison G., 2002, ApJL, in press {}{}
\bibitem{} Lilly S.J., Tresse L., Hammer F., Crampton D., Le Fevre O., 1995, ApJ, 455, L108{}{}
\bibitem{} Lilly S.J., Eales S.A., Gear W.K.P., Hammer F., Le Fevre O., et al., 1999, 518, 641{}{}
\bibitem{} Lutz D., Dunlop J.S., Almaini O., Andreani P., Blain A., et
al., 2001, A\&A, 378, L70{}{}
\bibitem{} Mainieri  V., et al., 2002, A\&A, 393, 425{}{}
\bibitem{} Mannucci F., Pozzetti L., Thompson D., Oliva E., Baffa C., et al., 2002, MNRAS, 329, L57{}{}
\bibitem{} Martini P., 2001, AJ, 121, 598{}{}
\bibitem{} McCarthy P.J., Carlberg R.G., Chen H.-W., Marzke R.O., et al., 2001, ApJ, 560, L131{}{}
\bibitem{} Mirabel I.F., Sanders D.B., 1989, ApJ, 340, L53{}{}
\bibitem{} Mobasher B., Cram L., Georgakakis A., Hopkins A., 1999, MNRAS, 308, 45{}{}
\bibitem{} Mohan N.R., Cimatti A., Rottgering H.J.A., Andreani P., et al., 2002, A\&A, in press{}{}
\bibitem{} Moriondo G., Cimatti A., Daddi E., 2000, A\& A, 364, 26{}{}
\bibitem{} Moran E.C., Lehnert M.D., Helfand D.J., 1999, ApJ, 526, 649{}{}
\bibitem{} Morrison G.E., 1999, PhD Thesis, University of New Mexico{}{}
\bibitem{} Morrison G.E., Owen F.N., Smail I., Dressler A., et al., 2002, ApJ, in prep{}{}
\bibitem{} Muxlow T.W.B., Wilkinson P.N., Richards A.M.S., Kellermann K.I., et al., 
           1999, NewAR, 43, 623{}{}
\bibitem{} Oliver S., Gruppioni C., Serjeant S., 1998, MNRAS, submitted{}{}
\bibitem{} Owen F.N., et al., 2002, in prep{}{}
\bibitem{} Packham C., Thompson K., Knapen J.H., Zurita A., Smail I., Greimel R., et al.,
           MNRAS, 2002, submitted{}{}
\bibitem{} Page M.J., Stevens J.A., Mittaz J.P.D., Carrera F.J., 2002, Science, 294, 2516{}{}
\bibitem{} Persson S.E., McCarthy P., Dressler A., Matthews
           K., 1993, in The Evolution of Galaixes and their Environments, eds., Shull M.,
           Thronson, H., NASA Tech.\ Report, GPO, p78{}{}
\bibitem{} Pierre M., Lidman C., Hunstead R., Alloin D., Casali M.,
           Cesarsky C., Chanial P., Duc P.-A., Fadda D., Flores H., Madden S.,
           Vigroux L., 2001, A\&A, 372, L45{}{}
\bibitem{} Poggianti B.M., Wu H., 2000, ApJ, 529, 157{}{}
\bibitem{} Pozzetti L., Mannucci F., 2000, MNRAS, 317 L17{}{}
\bibitem{} Ranalli P., Comastri A., Setti G., 2002, in New Visions of
           the X-ray Universe,  eds Jansen, F., (astro-ph/0202241)
\bibitem{} Richards E.A., 1998, PhD Thesis, U.\ of Virginia{}{}
\bibitem{} Richards E.A., 1999, ApJ, 513, L9{}{}
\bibitem{} Richards E.A., 2000, ApJ, 533, 611{}{}
\bibitem{} Richards E.A., Kellermann K.I., Fomalont E.B., Windhorst R.A., Partridge R.B.,
           1998, AJ, 116, 1039{}{}
\bibitem{} Richards E.A., Fomalont E.B., Kellermann K.I., Windhorst R.A., et al., 1999,
           ApJ, 562, L73{}{}
\bibitem{} Roche N., Lowenthal J., Koo D., 2002, MNRAS, in press{}{}
\bibitem{} Roche, N., Almaini, O., Dunlop, J.S., Ivison, R.J., Willott, C.,  2002,
	   MNRAS, submitted{}{}
\bibitem{} Sanders D.B., Mirabel I.F., 1996, ARA\&A, 34, 749{}{}
\bibitem{} Schindler S., Belloni P., Ikebe Y., Hattori M., Wambsganss J., Tanaka Y., 1998,
           A\&A, 338, 843{}{}
\bibitem{} Serjeant S., Mobasher B., Gruppioni C., Oliver S., 1999, MNRAS, 317, L29 {}{}
\bibitem{} Seitz C., Kneib J.-P., Schneider P., Seitz S., 1996, A\&A, 314, 707{}{}
\bibitem{} Smail I., Ivison R.J., Kneib J.-P., Cowie L.L.,
           Blain A.W., Barger A.J., Owen F.N., Morrison G., 1999a, MNRAS, 308, 1061{}{}
\bibitem{} Smail I., Morrison G., Gray, M.E., Owen F.N., Ivison
            R.J., Kneib J.-P., Ellis, R.S., 1999b, ApJ, 525, 609{}{}
\bibitem{} Smail I., Ivison R.J., Blain A.W., Kneib J.-P., 2002, MNRAS,
           331, 495{}{}
\bibitem{} Smail I., Ivison R.J., Gilbank D.G., Dunlop J.S., Keel W.C.,
           Stevens J.A., 2003, ApJ, in press{}{}
\bibitem{} Smith G.P., Treu T., Ellis R.S., Smail I., Kneib J.-P.,
           Frye B.L., 2001, ApJ, 562, 635{}{}
\bibitem{} Smith G.P., Smail I., Kneib J.-P., Czoske
           O., Ebeling H., Edge A.C., Pello R.,
           Ivison R.J., Packham C., Le Borgne J.-F., 2002a, MNRAS, 330, 1{}{}
\bibitem{} Smith G.P., Smail I., Kneib J.-P., Davis C.J., Takamiya M., Ebeling H., 
           Czoske O.,  2002b, MNRAS, in press{}{}
\bibitem{} Snigula J., Drory N., Bender R., Botzler C.S., Feulner G., Hopp U., 2002, MNRAS, in press{}{}
\bibitem{} Soifer B.T., Matthews K., Neugebauer G., Armus L.,
           Cohen J.G., Persson S.E., Smail I., 1999, AJ, 118, 2065{}{}
\bibitem{} Stanford S.A., Eisenhardt P.R.M., Dickinson M., 1995, ApJ, 450, 512{}{}
\bibitem{} Stevens, J.A., Page, M.J., Ivison, R.J., Smail, I., Lehmann I., Hassinger, G., Szokoly G.,
	   2002, MNRAS, submitted{}{}
\bibitem{} Thompson D., Beckwith S.V.W., Fockenbrock R., Fried J.,
           Hippelein H., Huang J.-S., von Kuhlmann B., Leinhert C.,
           Meisenheimer K., Phleps S., Roeser H.-J., Thommes E.,
           Wolf C., 1999, ApJ, 523, 100{}{}
\bibitem{} Trager S.C., Faber S.M., Dressler A., Oemler, A., 1997, ApJ, 485, 92{}{}
\bibitem{} Trentham N., Kormendy J., Sanders D., 1999, AJ, 117, 1152{}{}
\bibitem{} Waddington I., Windhorst R.A., Cohen S.H., Partridge R.B., Spinrad H., Stern D.,
           1999, ApJ, 526, L77{}{}
\bibitem{} Waddington I., Windhorst R.A., Dunlop J.S., Koo D.C., Peacock J.A, 2000, MNRAS, 317, 801{}{}
\bibitem{} Willott C.J., Rawlings S., Blundell K.M., 2001, MNRAS, 324, 1{}{}
\bibitem{} Windhorst R.A., Gordon J.M., Pascarelle S.M., Schmidtke P.C., Keel W.C.,
           Burkey J.M., Dunlop J.S, 1994, 435, 577{}{}
\bibitem{} Yan L., McCarthy P.J., Freudling W., et al., 1999, ApJ, 519, 47{}{}
\bibitem{} Yan L., 2001, in Gas \& Galaxy Evolution, Hibbard J.E., Rupen M.P.,
           van Gorkom J.H., ASP Conf.\ Series, 240, p117{}{}
\bibitem{} Yun, M.S., Reddy, N.A., Condon, J.J., 2001, ApJ, 554,
	   803{}{}
\end{thebibliography}
\end{document}